\pdfoutput=1
\RequirePackage{fix-cm}
\documentclass[twocolumn]{svjour3}          
\smartqed  
\usepackage{amsmath,amssymb,array}
\usepackage{graphicx}
\usepackage{hyperref}
\usepackage{booktabs}
\usepackage{multirow}
\usepackage[authoryear]{natbib}
\usepackage{verbatim}
\usepackage{enumitem}

\allowdisplaybreaks
\newcommand{\bs}{\boldsymbol}
\hypersetup{draft}

\begin{document}

\title{Clustering Multivariate Data using Factor Analytic Bayesian Mixtures with an Unknown Number of Components}

\titlerunning{Factor Analytic Bayesian Mixtures}        

\author{Panagiotis Papastamoulis
}


\institute{P. Papastamoulis \at
              Department of Statistics \\
                Athens University of Economics and Business\\
			  76 Patission Street
  10434 Athens\\
  Greece\\
              \email{\tt papastamoulis@aueb.gr}           
}

\date{Received: date / Accepted: date}

\maketitle

\begin{abstract}
Recent work on overfitting Bayesian mixtures of distributions offers a powerful framework for clustering multivariate data using a latent Gaussian model which resembles the factor analysis model. The flexibility provided by overfitting mixture models yields a simple and efficient way in order to estimate the unknown number of clusters and model parameters by Markov chain Monte Carlo (MCMC) sampling. The present study extends this approach by considering a set of eight parameterizations, giving rise to parsimonious representations of the covariance matrix per cluster. A Gibbs sampler combined with a prior parallel tempering scheme is implemented in order to approximately sample from the posterior distribution of the overfitting mixture. The  parameterization and number of factors is selected according to the Bayesian Information Criterion. Identifiability issues related to label  switching are dealt by post-processing the simulated output with the Equivalence Classes Representatives algorithm. The contributed method and software are demonstrated and compared to similar models estimated using the Expectation-Maximization algorithm on simulated and real datasets. The software is available online at \url{https://CRAN.R-project.org/package=fabMix}. 
\end{abstract}

\section{Introduction}
\label{intro}

Factor Analysis (FA) explains relationships among a set of observed variables using a set of latent variables. This is typically achieved by expressing the observed multivariate data as a linear combination of a smaller set of unobserved and uncorrelated variables known as factors. Let $\bs x = (\bs x_1,\ldots,\bs x_n)$ denote a random sample of $p$ dimensional observations with $\bs x_i\in\mathbb R^{p}$; $i= 1,\ldots,n$. Let $\mathcal N_p(\bs\mu,\bs\Sigma)$ denotes the $p$-dimensional normal distribution with mean $\bs\mu$ and covariance matrix $\bs\Sigma$ and also denote by $\bs{\mathrm{I}}_p$ the $p\times p$ identity matrix. The following equations summarize the typical FA model. 
\begin{eqnarray}
\label{eq:fa}
\bs x_i &=& \bs\mu + \bs\Lambda \bs y_i + \bs\varepsilon_i,\quad i = 1,\ldots,n\\
\label{eq:y}
(\bs y_i, \bs\varepsilon_i) &\sim& \mathcal N_q(\bs 0,\bs{\mathrm{I}}_q)\mathcal N_p(\bs 0,\bs\Sigma),\quad \mbox{iid for }i = 1,\ldots,n\\
\label{eq:diag}
\bs\Sigma &=& \mbox{diag}(\sigma_1^2, \ldots,\sigma_p^2)\\
\label{eq:x_given_y}
\bs x_i|\bs y_i &\sim& \mathcal N_p(\bs\mu+\bs\Lambda \bs y_i,\bs\Sigma),\quad \mbox{ind. for }i = 1,\ldots,n
\end{eqnarray}
Before proceeding note that we are not differentiating the notation between random variables and their corresponding realizations. Bold upper-case letters are used for matrices,  bold lower-case letters are used for vectors and normal text for scalars.

In Equation \eqref{eq:fa} we assume that $\bs x_i$ is expressed as a linear combination of a latent vector of factors $\bs y_i\in\mathbb R^{q}$. The $p\times q$ dimensional matrix $\bs\Lambda = (\lambda_{rj})$ contains the factor loadings, while $\bs\mu = (\mu_1,\ldots,\mu_p)$ contains the marginal mean of $\bs x_i$. The unobserved vector $\bs y_i$ lies on a lower dimensional space, that is, $q < p$ and it consists of uncorrelated features $y_{i1}, \ldots,y_{iq}$ as shown in Equation \eqref{eq:y}, where $\bs 0$
 denotes a vector of zeros. Note that the error terms $\bs\varepsilon_i$ are independent from $\bs y_i$. Furthermore, the errors are consisting of independent random variables $\varepsilon_{i1},\ldots,\varepsilon_{ip}$, as implied by the diagonal covariance matrix $\bs\Sigma$ in Equation \eqref{eq:diag}. As shown in Equation \eqref{eq:x_given_y}, the knowledge of the missing data 
($\bs y_i$) implies that the conditional distribution of $\bs x_i$ has a diagonal covariance matrix. The previous assumptions lead to 
\begin{equation}
\label{eq:x_marginal}
\bs x_i \sim \mathcal N_p(\bs\mu,\bs\Lambda\bs\Lambda^T + \bs\Sigma), \quad \mbox{iid for }i = 1,\ldots,n.
\end{equation}
According to Equation \eqref{eq:x_marginal}, the covariance matrix of the marginal distribution of $\bs x_i$ is equal to $\bs\Lambda\bs\Lambda^T + \bs\Sigma$. This is the crucial characteristic of factor analytic models, where they aim to explain high-dimensional dependencies using a set of lower-dimensional uncorrelated factors \citep{kim1978factor, bartholomew2011latent}.

Mixtures of Factor Analyzers (MFA) are generalizations of the typical FA model, by assuming that Equation \eqref{eq:x_marginal} becomes
\begin{equation}
\bs x_i \sim \sum_{k = 1}^{K}w_k\mathcal N_p(\bs\mu_{k},\bs\Lambda_{k}\bs\Lambda_{k}^T + \bs\Sigma_{k}), \mbox{ iid } i = 1,\ldots,n\label{eq:mixture}
\end{equation}
where $K$ denotes the number of mixture components. The vector of mixing proportions $\bs w := (w_1,\ldots,w_K)$ contains the weight of each component, with $0\leqslant w_k\leqslant 1$; $k = 1,\ldots,K$ and $\sum_{k=1}^{K}w_k = 1$. Note that the mixture components are characterized by different parameters $\bs\mu_k,\bs\Lambda_k,\bs\Sigma_k$, $k = 1,\ldots,K$. Thus, MFAs are particularly useful when the observed data exhibits unusual characteristics such as heterogeneity. That being said, this approach aims to capture the behaviour of each cluster within a component of the mixture model. A comprehensive perspective on the history and development of MFA models is given in Chapter 3 of the monograph by \cite{mcnicholas2016mixture}. 

Early works applying the Expectation-Maximization (EM) algorithm \citep{Dempster:77} for estimating MFA are the ones from \cite{ghahramani1996algorithm, tipping1999mixtures, McLachlan:00}. \cite{McNicholas2008, doi:10.1093/bioinformatics/btq498} introduced the family of parsimonious Gaussian mixture models (PGMM) by considering the case where the factor loadings and/or error variance may be shared or not between the mixture components. These models are estimated by the alternating expectation-conditional maximization algorithm \citep{RSSB:RSSB082} and have superior performance compared to other approaches \citep{McNicholas2008}.
 Under a Bayesian setup, \cite{Fokoue2003} estimate the number of mixture components and factors by simulating a continuous-time stochastic birth-death point process using a Birth-Death MCMC algorithm \citep{stephens2000}. More recently, \cite{PAPASTAMOULIS2018220} estimated Bayesian MFA models with an unknown number of components using overfitting mixtures. 

In recent years there is a growing progress on the usage of overfitting mixture models in Bayesian analysis \citep{rousseau2011asymptotic, overfitting}. An overfitting mixture model consists of a number of components  which is much larger than its true (and unknown) value. Under suitable prior assumptions (see Appendix A) introduced by \cite{rousseau2011asymptotic}, it has been shown that asymptotically the redundant components will have zero posterior weight and force the posterior distribution to put all its mass in the sparsest way to approximate the true density. Therefore, the inference on the number of mixture components can be based on the posterior distribution of the ``alive'' components of the overfitted model, that is, the components which contain at least one allocated observation. 

Other Bayesian approaches to estimate the number of components in a mixture model include the Reversible jump MCMC (RJMCMC) \citep{Green:95, Richardson:97,dellaportas2006multivariate, papRJ}, Birth-death 
MCMC (BDMCMC) \citep{stephens2000} and allocation sampling \citep{Nobile2007, papastamoulis2016bayesbinmix} algorithms. However, overfitting mixture models are straightforward to implement, while the rest approaches require either careful design of various move types that bridge models with different number of clusters, or analytical integration of parameters.

The overall message is that there is a need for developing an efficient Bayesian method that will combine the previously mentioned frequentist advances on parsimonious representations of MFAs and the flexibility provided by the Bayesian viewpoint. This study aims at filling this gap by extending the Bayesian method of \cite{PAPASTAMOULIS2018220} to the family of parsimonious Gaussian mixtures of \cite{McNicholas2008}. Furthermore, we illustrate the proposed method using the  {\tt R} \citep{ihaka:1996, R} package {\tt fabMix} \citep{fabMix} available as a contributed package from the Comprehensive R Archive Network at \url{https://CRAN.R-project.org/package=fabMix}. The proposed method efficiently deals with many inferential problems (see e.g.~\cite{doi:10.1080/01621459.2000.10474285}) related to mixture posterior distributions, such as (i) inferring the number of non-empty clusters using overfitting models, (ii) efficient exploration of the posterior surface by running parallel heated chains and (iii) incorporating advanced techniques that succesfully deal  with the label switching issue \citep{papastamoulis2016label}. 

The rest of the paper is organized as follows. Section \ref{sec:model} reviews the basic concepts of parsimonious MFAs. Identifiability problems and corresponding treatments are detailed in Section \ref{sec:labelSwitching}. The Bayesian model is introduced in Section \ref{sec:prior}. Section \ref{sec:inference} presents the full conditional posterior distributions of the model. The MCMC algorithm is described in Section \ref{sec:other}. A detailed presentation of the main function of the contributed {\tt R} package is given in Section \ref{sec:fabMixFunction}. Our method is illustrated and compared to similar models estimated by the EM algorithm in Sections \ref{sec:sim} and \ref{sec:yeast} using an extended simulation study and 4 publicly available datasets, respectively. We conclude in Section \ref{sec:summary} with a summary of our findings and directions for further research. An Appendix contains further discussion on overfitting mixture models (Appendix A), details of the MCMC sampler (Appendix B) and additional simulation results (Appendix C).

\section{Parsimonious Mixtures of Factor Analyzers}\label{sec:model}

Consider the latent allocation variables $z_i$ which assign observation $\bs x_i$ to a component $k =1,\ldots,K$ for $i = 1,\ldots,n$. A-priori each observation is generated from component $k$ with probability equal to $w_k$, that is,
\begin{equation}\label{eq:z_prior}
\mathrm{P}(z_i = k) = w_k,\quad k = 1,\ldots,K,
\end{equation}
independent for $i = 1,\ldots,n$.  Note that the allocation vector $\bs z := (z_1,\ldots,z_n)$ is not observed, so it should be treated as missing data. We assume that $\bs z_i$ and $\bs y_i$ are independent, thus Equation \eqref{eq:y} is now written as: 
\begin{equation}
(\bs y_i, \bs\varepsilon_i|z_i = k)\sim \mathcal N_q(\bs 0,\bs{\mathrm{I}}_q)\mathcal N_p(\bs 0,\bs\Sigma_{k}), \label{eq:varepsilon}
\end{equation}
and conditional on the cluster membership and latent factors we obtain that
\begin{equation}
(\bs x_i|z_i = k,\bs y_i) \sim \mathcal N_p(\bs\mu_{k}+\bs\Lambda_{k} \bs y_i,\bs\Sigma_{k}).\label{eq:fullConditionalX}
\end{equation}
Consequently,
\begin{equation}
(\bs x_i|z_i = k) \sim \mathcal N_p(\bs\mu_{k},\bs\Lambda_{k}\bs\Lambda_{k}^{T}+\bs\Sigma_{k}),\label{eq:ConditionalX}
\end{equation}
independent for $i=1,\ldots,n$. From Equations \eqref{eq:z_prior} and \eqref{eq:ConditionalX} we derive that the marginal distribution of $\bs x_i$ is the finite mixture model in Equation \eqref{eq:mixture}. 

Following \cite{McNicholas2008}, the factor loadings and error variance per component may be common or 
not among the $K$ components in Equation  \eqref{eq:mixture}. If the factor are constrained, then: \begin{equation}\label{eq:lambda_common}\bs \Lambda_1=\ldots=\bs\Lambda_K = \bs\Lambda.\end{equation} If the error variance is constrained, then: \begin{equation}\label{eq:sigma_common}\bs \Sigma_1=\ldots=\bs\Sigma_K = \bs\Sigma.\end{equation} Furthermore, the error variance may be isotropic (i.e.~proportional to the identity matrix) or not and depending on whether contstraint \eqref{eq:sigma_common} is disabled or enabled: 
\begin{eqnarray}
\label{eq:iso1}
\bs \Sigma_k &=& \sigma^2_k \bs{\mathrm{I}}_p; k = 1,\ldots,K \quad \mbox{ or}\\
\label{eq:iso2}
\bs \Sigma_k &=& \sigma^2 \bs{\mathrm{I}}_p; k = 1,\ldots,K .
\end{eqnarray} 
We note that under constraint \eqref{eq:iso1},  the model is referred to as a mixture of probabilistic principal component analyzers \citep{tipping1999mixtures}.

Depending on whether a particular constraint is present or not, the following set of 8 parameterizations arises. 
\begin{eqnarray*}
\mbox{UUU:} && \boldsymbol{x}_i \sim \sum\nolimits_{k = 1}^{K}w_k\mathcal N_p(\bs\mu_{k},\bs\Lambda_{k}\bs\Lambda_{k}^T + \bs\Sigma_{k})\\
\mbox{UCU:} && \boldsymbol{x}_i \sim \sum\nolimits_{k = 1}^{K}w_k\mathcal N_p(\bs\mu_{k},\bs\Lambda_{k}\bs\Lambda_{k}^T + \bs\Sigma)\\
\mbox{UUC:} && \boldsymbol{x}_i \sim \sum\nolimits_{k = 1}^{K}w_k\mathcal N_p(\bs\mu_{k},\bs\Lambda_{k}\bs\Lambda_{k}^T + \sigma^2_k\bs{\mathrm{I}}_p)\\
\mbox{UCC:} && \boldsymbol{x}_i \sim \sum\nolimits_{k = 1}^{K}w_k\mathcal N_p(\bs\mu_{k},\bs\Lambda_{k}\bs\Lambda_{k}^T + \sigma^2\bs{\mathrm{I}}_p)\\
\mbox{CUU:} && \boldsymbol{x}_i \sim \sum\nolimits_{k = 1}^{K}w_k\mathcal N_p(\bs\mu_{k},\bs\Lambda\bs\Lambda^T + \bs\Sigma_{k})\\
\mbox{CCU:} && \boldsymbol{x}_i \sim \sum\nolimits_{k = 1}^{K}w_k\mathcal N_p(\bs\mu_{k},\bs\Lambda\bs\Lambda^T + \bs\Sigma)\\
\mbox{CUC:} && \boldsymbol{x}_i \sim \sum\nolimits_{k = 1}^{K}w_k\mathcal N_p(\bs\mu_{k},\bs\Lambda\bs\Lambda^T + \sigma^2_k\bs{\mathrm{I}}_p)\\
\mbox{CCC:} && \boldsymbol{x}_i \sim \sum\nolimits_{k = 1}^{K}w_k\mathcal N_p(\bs\mu_{k},\bs\Lambda\bs\Lambda^T + \sigma^2\bs{\mathrm{I}}_p)\\
\end{eqnarray*}
independent for $i = 1,\ldots,n$. Following the {\tt pgmm} nomenclature \citep{McNicholas2008}: the first, second and third letter denotes whether $\bs\Lambda_k$, $\bs\Sigma_k=\mathrm{diag}(\sigma^2_{k1},\\ \ldots,\sigma^2_{kp})$ and $\sigma^2_{kj}$, $k=1,\ldots,K$; $j=1,\ldots,p$, are constrained (C)  or unconstrained (U), respectively. A novelty of the present study is to offer a  Bayesian framework for estimating the whole family of the previous parameterizations (note that \cite{PAPASTAMOULIS2018220} estimated the UUU and UCU parameterizations).

\subsection{Label switching and other identifiability problems}\label{sec:labelSwitching}

Let $L(\bs w, \bs\theta, \bs\phi|\bs x) = \prod_{i=1}^{n}\sum_{k=1}^{K}w_kf(x_i|\theta_k, \bs\phi)$, $(\bs w, \bs\theta,\bs\phi)\in\mathcal P_{K-1}\times\Theta^{K}\times\Phi$ denote the likelihood function of a mixture of $K$ densities, where $\mathcal P_{K-1}$ denotes the parameter space of the mixing proportions $\bs w$, $\bs\theta = (\bs\theta_1,\ldots,\bs\theta_K)$ are the component-specific parameters and $\bs\phi$ denotes a (possibly empty) collection of parameters that are common between all components. For instance, consider the UCU parameterization where $\bs\theta_k = (\bs\mu_k, \bs\Lambda_k)$  for $k = 1,\ldots,K$ and $\bs\phi = \bs\Sigma$. For any permutation $\tau=(\tau_1,\ldots,\tau_K)$ of the set $\{1,\ldots,K\}$, the likelihood of mixture models is invariant to permutations of the component labels: $L(\bs w, \bs\theta, \bs\phi|\bs x) = L(\tau \bs w, \tau\bs\theta, \bs\phi|\bs x)$.
Thus, the likelihood surface of a mixture model with $K$ components will exhibit $K!$ symmetric areas. If $(\bs w^{*},\bs\theta^{*},\bs\phi^{*})$ corresponds to a mode of the likelihood, the same will hold for any permutation $(\tau\bs w^{*}, \tau\bs\theta^{*},\bs\phi^{*})$. 

Label switching \citep{redner1984mixture} is the commonly used term to describe this phenomenon. Under a Bayesian point of view, in the case that the prior distribution is also invariant to permutations (which is typically the case, see e.g.~\cite{Marin:05, Papastamoulis2013}), the same invariance property will also hold for the posterior distribution $f(\bs w, \bs\theta, \bs \phi|\bs x)$. Consequently, the marginal posterior distributions of mixing proportions and component-specific parameters will be coinciding, i.e.: $f(w_1|\bs x) = \ldots = f(w_K|\bs x)$ and $f(\theta_1|\bs x) = \ldots = f(\theta_K|\bs x)$. Thus, when approximating the posterior distribution  via MCMC sampling, the standard practice of ergodic averages for estimating quantities of interest (such as the  mean of the marginal posterior distribution for each parameter) becomes meaningless. In order to deal with this identifiability problem we post-process the simulated MCMC output using a deterministic relabelling algorithm, that is, the Equivalence Classes Representatives (ECR) algorithm \citep{Papastamoulis:10, papastamoulis2014handling}, as implemented in the {\tt R} package {\tt label.switching} \citep{papastamoulis2016label}. 

A second source of identifiability problems is related to  orthogonal transformations of the matrix of factor loadings.  A popular practice \citep{doi:10.1093/rfs/9.2.557, Fokoue2003, doi:10.1111/bmsp.12019, PAPASTAMOULIS2018220} to overcome this issue, is to preassign values to some entries of $\bs\Lambda$, in particular we set the entries of the upper diagonal of the first  $q\times q$ block matrix of $\bs\Lambda$ equal to zero:
$$\bs\Lambda = \begin{pmatrix}
\lambda_{11} & 0 & \cdots & 0\\
\lambda_{21} & \lambda_{22} & \cdots & 0\\
\vdots & \vdots & \ddots & \vdots\\
\lambda_{q1} & \lambda_{q2} & \cdots & \lambda_{qq}\\
\vdots & \vdots & \ddots & \vdots\\
\lambda_{p1} & \lambda_{p2} & \cdots & \lambda_{pq}
\end{pmatrix}.$$

Another problem is related to the so-called ``sign switching'' phenomenon, see e.g.~\cite{conti2014bayesian}. Simultaneously switching the signs of a given row $r$ of $\bs\Lambda$; $r = 1,\ldots,p$ and $\bs y_i$ does not alter the likelihood. Thus, $\bs\Lambda$ and $\bs y_i$; $i = 1, \ldots,n$ are not marginally identifiable due to sign-switching across the MCMC trace. However, this is not a problem in our implementation, since all parameters of the marginal density of $\bs x_i$ in \eqref{eq:mixture} are identified (see also the discussion for sign-invariant parametric functions in \cite{PAPASTAMOULIS2018220}). 

Parameter expanded approaches are preferred in the recent literature \citep{bhattacharya2011sparse, mcparland2017clustering}, because the mixing of the MCMC sampler is improved. In our implementation, we are able to obtain excellent mixing using the popular approach of restricting elements of $\bs \Lambda$: the reader is referred to Figure 2 of \citet{PAPASTAMOULIS2018220}, where it is obvious that our MCMC sampler has the ability to rapidly move between the multiple modes of the target posterior distribution of $\bs\Lambda$ (more details on convergence diagnostics are also presented in Appendix A.4 of \citet{PAPASTAMOULIS2018220}).

\subsection{Prior assumptions}\label{sec:prior}

We assume that the number of mixture components ($K$) has a sufficiently large value so that it overestimates the ``true'' number of clusters. Unless otherwise stated, the default choice is $K = 20$. All prior assumptions of the overfitting mixture models are discussed in detail in \cite{PAPASTAMOULIS2018220}. For ease of presentation we repeat them in this section. Let $\mathcal D(\cdots)$  denote the Dirichlet distribution and $\mathcal G(\alpha,\beta)$ denote the Gamma distribution with mean $\alpha/\beta$. Let also $\bs\Lambda_{kr\cdot}$ denote the $r$-th row of the matrix of factor loadings $\bs\Lambda_k$; $k = 1,\ldots,K$; $r = 1,\ldots,p$. The following prior assumptions are imposed on the model parameters:

\begin{eqnarray}
\label{eq:dirichlet_prior}
\bs w &\sim&\mathcal D\left(\gamma,\ldots,\gamma\right), \quad \gamma = \frac{1}{K} \\
\bs\mu_k &\sim&\mathcal N_p(\bs\xi, \bs\Psi), \label{eq:mu_prior} \quad\mbox{iid for }k = 1,\ldots,K\\
\bs\Lambda_{kr\cdot} &\sim&\mathcal N_{\nu_r}(\bs 0,\bs\Omega), \quad\mbox{iid. for }r = 1,\ldots,p \label{eq:lambda_prior} \\
\sigma_{kr}^{-2} &\sim& \mathcal G(\alpha,\beta), \quad\mbox{iid for }k = 1,\ldots,K; r = 1,\ldots,p \label{eq:sigma_prior}\\
\omega_{\ell}^{-2} &\sim& \mathcal G(g,h), \quad\mbox{iid for }\ell = 1,\ldots,q \label{eq:omega_prior}
\end{eqnarray}
where all variables are assumed mutually independent and $\nu_r =\min\{r,q\}$; $r=1,\ldots,p$; $\ell = 1,\ldots,q$; $j=1,\ldots,K$.  In Equation \eqref{eq:lambda_prior}  $\bs\Omega = \mbox{diag}(\omega_1^2,\ldots,\omega_q^2)$ denotes a $q\times q$ diagonal matrix, where the diagonal entries are distributed independently according to Equation \eqref{eq:omega_prior}. A graphical representation of the hierarchical model is given in Figure 1 of \cite{PAPASTAMOULIS2018220}. The default values of the remaining fixed hyper-parameters are given in  Appendix B.

The previous assumptions refer to the case of the unconstrained parameter space, that is, the UUU paramaterization. Clearly, they should be modified accordingly when a constrained model is used. Under constraint \eqref{eq:lambda_common}, the prior distribution in Equation \eqref{eq:lambda_prior} becomes $\bs\Lambda_{r\cdot} \sim\mathcal N_{\nu_r}(\bs 0,\bs\Omega)$, independent for $r = 1,\ldots,p$. Under constraints \eqref{eq:sigma_common} and \eqref{eq:iso1}, the prior distribution in Equation \eqref{eq:sigma_prior} becomes $\sigma_{r}^{-2} \sim \mathcal G(\alpha,\beta)$, independent for $r = 1,\ldots,p$. Finally, under constraints \eqref{eq:sigma_common} and \eqref{eq:iso2}, the prior distribution in Equation \eqref{eq:sigma_prior} becomes $\sigma^{-2} \sim \mathcal G(\alpha,\beta)$.

\section{Inference}\label{sec:inference}

This section describes the full conditional posterior distributions of model parameters and the corresponding MCMC sampler. Due to conjugacy, all full conditional posterior distributions are available in closed forms. 

\subsection{Full conditional posterior distributions}\label{sec:sampler}
Let us define the following quantities:
\begin{eqnarray*}
n_k &=& \sum_{k=1}^{K}I(z_i = k)\\
\bs A_k &=& n_k\bs\Sigma_k^{-1} + \Psi^{-1} \\
\bs B_k &=& \bs\Sigma_k^{-1}\sum_{k=1}^{K}I(z_i=k)\left(\bs x_i - \bs\Lambda_k\bs y_i\right) + \bs\xi\bs\Psi^{-1}\\
\bs\tau_{kr} &=& \frac{\sum_{i=1}^{n}I(z_i = k)(x_{ir} - \mu_{kr})\bs y_i^{T}}{\sigma_{kr}^2}\\
\bs\Delta_{kr} &=& \frac{\sum_{i=1}^{n}I(z_i = k)\bs y_i\bs y_i^{T}}{\bs\sigma_{kr}^2}\\
s_{kr} &=& \sum_{i=1}^{n}I(z_i = k)\left( x_{ir} - \mu_{kr} - \bs\Lambda_{kr\cdot}\bs y_i\right)^2\\
\bs T &=& \sum_{k=1}^{K}\sum_{r=1}^{p}\bs\Lambda_{kr\cdot}\bs\Lambda_{kr\cdot}^{T}\\
\bs M_k & = & \bs{\mathrm{I}}_q + \bs\Lambda^{T}_k\bs\Sigma^{-1}_k\bs\Lambda^{T}_k
\end{eqnarray*}
for $k = 1,\ldots,K$; $r = 1,\ldots,p$. For a generic sequence of the form $\{G_{rc}; r \in \mathcal R, c\in \mathcal C\}$ we also define $G_{\bullet c}=\sum_{r}G_{rc}$ and $G_{r\bullet}=\sum_{c}G_{rc}$. Finally, $(x|\cdots)$ denotes the conditional distribution of $x$ given the value of all remaining variables.

From Equations \eqref{eq:mixture} and                             \eqref{eq:z_prior} it immediately follows that for $k=1,\ldots,K$
\begin{equation}\label{eq:z_fc}
\mathrm{P}(z_i = k|\cdots) \propto w_kf\left(\bs x_i;\bs\mu_k,\bs\Lambda_k\bs\Lambda_k^{T} + \bs\Sigma_k\right), 
\end{equation}
independent for $i  =1,\ldots,n$, where $f(\cdot;\bs\mu,\bs\Sigma)$ denotes the probability density function of the multivariate normal distribution with mean $\bs\mu$ and covariance matrix $\bs\Sigma$. Note that in order to compute the right hand side of the last equation, inversion of the $p\times p$ matrix $\bs\Lambda_k\bs\Lambda_k^{T} + \bs\Sigma_k$ is required. Using the Sherman--Morrison--Woodbury formula (see e.g.~\cite{10.2307/2030425}), the inverse matrix is equal to $\bs\Sigma_k^{-1} - \bs\Sigma_k^{-1}\bs\Lambda_k \bs M_{k}^{-1}\bs\Lambda_k^T\bs\Sigma^{-1}_k$, 
for $k = 1,\ldots,K$. The full conditional posterior distribution of mixing proportions is a Dirichlet distribution with parameters
\begin{equation}\label{eq:w_fc}
\bs w|\cdots \sim \mathcal D(\gamma + n_1,\ldots,\gamma+n_K). 
\end{equation}

The full conditional posterior distribution of the marginal mean per component is
\begin{equation}
\label{eq:mu_fc}
\bs\mu_k|\cdots \sim \mathcal N_p\left( \bs A_k^{-1}\bs B_k, \bs A_k^{-1} \right), 
\end{equation}
independent for $k = 1\ldots,K$.

The full conditional posterior distribution of the factor loadings without any restriction is 
\begin{equation}
\label{eq:lambda_fc}
\bs\Lambda_{kr\cdot}|\cdots \sim \mathcal N_{\nu_r}\left(\left[\bs\Omega^{-1} + \bs\Delta_{kr}\right]^{-1}\bs\tau_{kr},\left[\bs\Omega^{-1} + \bs\Delta_{kr}\right]^{-1}\right),
\end{equation}
independent for $k = 1,\ldots,K$; $r = 1,\ldots,p$. Under constraint \eqref{eq:lambda_common} we obtain that
\begin{equation}
\label{eq:lambda_fc2}
\bs\Lambda_{r\cdot}|\cdots \sim \mathcal N_{\nu_r}\left(\left[\bs\Omega^{-1} + \bs\Delta_{\bullet r}\right]^{-1}\bs\tau_{\bullet r},\left[\bs\Omega^{-1} + \bs\Delta_{\bullet r}\right]^{-1}\right),
\end{equation}
independent for $r = 1,\ldots,p$. 

The full conditional distribution of error variance without any restriction is
\begin{equation}
\label{eq:sigma_fc}
\sigma_{kr}^{-2}|\cdots \sim \mathcal G\left(\alpha + n_k/2,\beta + s_{kr}/2\right),
\end{equation}
independent for $k = 1,\ldots,K$; $r = 1,\ldots,p$. Under constraint \eqref{eq:sigma_common} we obtain that 
\begin{equation}
\label{eq:sigma_fc0}
\sigma_{r}^{-2}|\cdots \sim \mathcal G(\alpha + n/2,\beta + s_{\bullet r}/2),
\end{equation}
independent for $r = 1,\ldots,p$. Under constraints \eqref{eq:sigma_common} and \eqref{eq:iso1} we obtain that
\begin{equation}
\label{eq:sigma_fc1}
\sigma_{k}^{-2}|\cdots \sim \mathcal G(\alpha + n_kp/2,\beta + s_{k\bullet}/2),
\end{equation}
independent for $k = 1,\ldots,K$. Under constraints \eqref{eq:sigma_common} and \eqref{eq:iso2} we obtain that
\begin{equation}\label{eq:sigma_fc2}
\sigma^{-2}|\cdots \sim \mathcal G(\alpha + np/2,\beta + s_{\bullet\bullet}/2).
\end{equation}
The full conditional distribution of latent factors is given by
\begin{equation}
\label{eq:y_fc}
\bs y_i|\cdots \sim\mathcal N_q\left(\bs M_{z_i}^{-1}\bs\Lambda_{z_i}^{T}\bs\Sigma^{-1}_{z_i}(\bs x_i - \bs\mu_{z_i}), 
\bs M_{z_i}^{-1}
\right),
\end{equation}
independent for $i = 1,\ldots,n$. Finally, the full conditional distribution for $\omega_\ell$ is
\begin{equation}
\label{eq:omega_fc}
\omega_\ell^{-2}|\cdots\sim\mathcal G\left(g + Kp/2, h + T_{\ell\ell}/2\right),
\end{equation}
while under constraint \eqref{eq:lambda_common} we obtain that
\begin{equation}
\label{eq:omega_fc_constrained}
\omega_\ell^{-2}|\cdots\sim\mathcal G\left(g + p/2, h + T_{\ell\ell}/2K\right),
\end{equation}
independent for $\ell = 1,\ldots,q$.

\subsection{MCMC sampler}\label{sec:other}

Given the number of factors ($q$) and a model parameterization, a Gibbs sampler \citep{geman, gelfand} coupled with a prior parallel tempering scheme \citep{geyer1991, geyer1995, Altekar12022004} is used in order to produce a MCMC sample from the joint posterior distribution. Each heated chain ($j = 1,\ldots,\mbox{\tt nChains}$) corresponds to a model with identical likelihood as the original, but with a different prior distribution. Although the prior tempering can be imposed on any subset of parameters, it is only applied to the Dirichlet prior distribution of mixing proportions \citep{overfitting}. 
The inference is based on the output of the first chain  ($j = 1$) of the prior parallel tempering scheme \citep{overfitting}. The number of factors and model parameterization is selected according to the Bayesian Information Criterion (BIC) \citep{Schwarz:78}, conditional on the most probable number of alive clusters per model (see \cite{PAPASTAMOULIS2018220} for a detailed comparison of BIC with other alternatives).

Let $\mathcal M$ and $\mathcal Q$ denote the set of model parameterizations and number of factors. In the following pseudocode, $x\leftarrow [y|z]$ denotes that $x$ is updated from a draw from the distribution $f(y|z)$ and $\theta_j^{(t)}$ denotes the value of $\theta$ at the $t$-th iteration of the $j$-th chain. 
\begin{enumerate}[leftmargin=0.4cm]
\item For $(m,q) \in \mathcal M\times \mathcal Q$
\begin{enumerate}[leftmargin=0.4cm]
\item Obtain initial values $(\bs\Omega_j^{(0)}$, $\bs\Lambda_{m;j}^{(0)}$, $\bs\mu^{(0)}_j$,  $\bs z^{(0)}_j$, $\bs\Sigma_{m;j}^{(0)}$, $\bs  w^{(0)}_j$, $\bs y^{(0)}_j)$ by running the overfitting initialization scheme, for $j = 1,\ldots,\mbox{\tt nChains}$.
\item For MCMC iteration $t = 1,2,\ldots$ update 
\begin{enumerate}[leftmargin=0.1cm]
\item For chain $j = 1,\ldots,\mbox{\tt nChains}$
\begin{enumerate}[leftmargin=0.2cm]
\item $\bs\Omega^{(t)}_j \leftarrow \left[\bs\Omega|\bs\Lambda_{mj}^{(t-1)}\right]$. \\If $m \in \{\mbox{UUU,UCU,UUC, UCC}\}$ use \eqref{eq:omega_fc}\\ else use \eqref{eq:omega_fc_constrained}.
\item $\bs\Lambda_{m;j}^{(t)}\leftarrow\left[\bs\Lambda|\bs\Omega_j^{(t)}, \bs \mu_j^{(t-1)},\bs\Sigma_{m;j}^{(t-1)}, \bs x, \bs y_j^{(t-1)}, \bs z_j^{(t-1)}\right]$\\ If $m \in \{\mbox{UUU,UCU,UUC, UCC}\}$ use \eqref{eq:lambda_fc}\\ else use \eqref{eq:lambda_fc2}.
\item $\bs\mu_j^{(t)}\leftarrow\left[\bs\mu|\bs\Lambda_m^{(t)},\bs \Sigma_m^{(t-1)}, \bs x, \bs y^{(t-1)}, \bs z_j^{(t-1)}\right]$ according to \eqref{eq:mu_fc}.
\item $\bs z_j^{(t)}\leftarrow \left[\bs z|\bs w_j^{(t-1)},\bs \mu_j^{(t)}, \bs\Lambda_{m;j}^{(t)}, \bs\Sigma_{m;j}^{(t-1)}, \bs x\right]$ according to \eqref{eq:z_fc}.
\item  $\bs w_j^{(t)}\leftarrow\left[\bs w|\bs z_j^{(t)}\right]$ according to \eqref{eq:w_fc} with prior parameter $\gamma = \gamma_{(j)}$.
\item $\bs\Sigma_{m;j}^{(t)}\leftarrow\left[\bs\Sigma|\bs x, \bs z_j^{(t)}, \bs\mu_j^{(t)}, \bs\Lambda_{m;j}^{(t)}, \bs y_j^{(t-1)}\right]$\\ If $m \in \{\mbox{UUU,CUU}\}$ use \eqref{eq:sigma_fc}\\ else if  $m \in \{\mbox{UCU,CCU}\}$  use \eqref{eq:sigma_fc0}\\ else if $m \in \{\mbox{UUC,CUC}\}$  use \eqref{eq:sigma_fc1}\\ else use \eqref{eq:sigma_fc2}. 
\item $\bs y_j^{(t)}\leftarrow\left[\bs y|\bs x, \bs z_j^{(t)}, \bs\mu_j^{(t)}, \bs\Sigma_{m;j}^{(t)},\bs \Lambda_{m;j}^{(t)}\right]$ according to \eqref{eq:y_fc}.
\end{enumerate}
\item Select randomly $1\leqslant j^*\leqslant \mbox{\tt nChains}-1$ and propose to swap the states of chains $j^*$ and $j^*+1$.
\end{enumerate}
\item For chain $j = 1$ compute BIC conditionally on the most probable number of alive clusters.
\end{enumerate}
\item Select the best $(m,q)$ model corresponding to chain $j = 1$ according to BIC and reorder the simulated output of the selected model according to ECR algorithm, conditional on the most probable number of alive clusters.
\end{enumerate}

The MCMC algorithm is initialized using random starting values arising from the ``overfitting initialization'' procedure introduced by \cite{PAPASTAMOULIS2018220}. For further details on steps 1.(a) (MCMC initialization) and 1.(b).ii (prior parallel tempering scheme) the reader is referred to Appendix B (see also Sections 2.6, 2.7 and 2.9 of \cite{PAPASTAMOULIS2018220}). 

\begin{table*}[t]
\caption{Arguments of the {\tt fabMix()} function.}
\label{tab:input}
\begin{tabular}{lp{12.6cm}}
\toprule
Argument & Description\\ \midrule
{\tt model} & Any non-empty subset of {\tt c("UUU", "CUU", "UCU", "CCU", "UCC", "UUC", "CUC", "CCC")},
          indicating the fitted models. By default, all models
          are fitted.\\
{\tt Kmax} & Number of components in the overfitted mixture (integer, at least equal to two). Default: 20.\\
{\tt nChains} & Number of parallel (heated) chains. When {\tt dirPriorAlphas} is supplied, this argument can be ignored.\\
{\tt dirPriorAlphas} &  vector of length {\tt nChains} in the form of an increasing
          sequence of positive scalars. Each entry contains the
          (common) prior Dirichlet parameter for the corresponding
          chain. Default: {\tt dirPriorAlphas = c(1, 1 + dN*(2:nChains -
          1))/Kmax}, where {\tt dN = 1}, for {\tt nChains > 1}. Otherwise,
          {\tt dirPriorAlphas = 1/Kmax}.\\
{\tt rawData} & The observed data in the form of an $n\times p$ matrix. Clustering is performed on the rows of the matrix.\\
{\tt outDir} & Name of the output folder. An error is thrown if the
          directory already exists inside the current working
          directory. Note: it should not correspond to an absolute
          path, e.g.: {\tt outDir = `example`} is acceptable, but
          {\tt outDir = `C:\textbackslash User\textbackslash Documents\textbackslash example`} is not.\\
{\tt mCycles} & Number of MCMC cycles. Each cycle consists of {\tt nIterPerCycle}
          MCMC iterations. At the end of each cycle a swap of the state
          of two randomly chosen adjacent chains is attempted.\\
{\tt burnCycles} & Number of cycles that will be discarded as burn-in period.\\
{\tt g} & Prior parameter $g$. Default value: {\tt g = 0.5}.\\
{\tt h} & Prior parameter $h$. Default value: {\tt g = 0.5}.\\
{\tt alpha\_sigma} & Prior parameter $\alpha$. Default value: {\tt alpha\_sigma = 0.5}.\\
{\tt beta\_sigma} & Prior parameter $\beta$. Default value: {\tt beta\_sigma = 0.5}.\\
{\tt q} & A vector of strictly positive integers, containing the number of factors to be fitted.\\
{\tt normalize} & Logical value indicating whether the observed data will be normalized. Default value: TRUE (recommended)\\
{\tt nIterPerCycle} & Number of iteration per MCMC cycle. Default value: 10.\\
{\tt warm\_up\_overfitting} & Number of iterations for the overfitting
          initialization scheme. Default value: 500.\\
{\tt warm\_up}& Number of iterations that will be used to initialize the
          models before starting proposing switchings. Default value:
          5000.\\
{\tt overfittingInitialization} & Logical value indicating whether the chains
          are initialized via the overfitting initialization scheme.
          Default: TRUE (recommended). \\
{\tt rmDir} & Logical value indicating whether to delete the {\tt outDir}
          directory. Default: TRUE.\\
{\tt parallelModels} & Model-level parallelization: An optional integer
          specifying the number of cores that will be used in order to
          fit in parallel each member of {\tt model}. Default: {\tt NULL} (no
          model-level parallelization).\\
\bottomrule
\end{tabular}
\end{table*}

\begin{table*}[t]
\caption{Output returned to the user of the {\tt fabMix()} function.}
\label{tab:output}
\begin{tabular}{lp{12.6cm}}
\toprule
Object & Description\\ \midrule
       {\tt bic} &
                Bayesian Information Criterion per model and number of
          factors.\\
             
        {\tt class} & The estimated single best clustering of the observations
          according to the selected model.\\
        
        {\tt n\_Clusters\_per\_model} &
The most probable number of clusters (number of
          non-empty components of the overfitting mixture) per model and
          number of factors\\
        
        {\tt posterior\_probability} &
                The posterior probability of the estimated
          allocations according to the selected model\\
        
        {\tt covariance\_matrix}&
                The estimated posterior mean of the covariance
          matrix per cluster according to the selected model\\
        
        {\tt mu}&
               The estimated posterior mean of the mean per cluster
          according to the selected model\\
                {\tt weights} & The estimated posterior mean of the mixing proportions
          according to the selected model\\
        {\tt mcmc} &  A list containing the MCMC draws for the parameters of the
          selected model.\\
        {\tt Kmap\_prob} & The posterior probability of the Maximum A Posteriori number
          of alive clusters for each parameterization and factor level.\\
\bottomrule
\end{tabular}
\end{table*}

\section{Using the {\tt fabMix} package}\label{sec:fabMixFunction}

The main function of the {\tt fabMix} package is {\tt fabMix()}, with its arguments shown in Table \ref{tab:input}. This function takes as input a matrix {\tt rawData} of observed data where rows and columns correspond to observations and variables of the dataset, respectively. The parameters of the Dirichlet prior distribution ($\gamma_{(j)}; j = 1,\ldots,\mbox{nChains}$) of the mixing proportions are controlled by {\tt dirPriorAlphas}. The range for the number of factors is specified in the {\tt q} argument. Valid input for {\tt q} is any positive integer vector between 1 and the Leddermann bound \citep{ledermann1937rank} implied by the number of variables in the dataset. By default, all 8 parameterizations are fitted, however the user can specify in {\tt model} any non-empty subset of them.

The {\tt fabMix()} function simulates a total number of $\mbox{{\tt nChains}}\times\mbox{{\tt length(models)}}\times \mbox{{\tt length(q)}}$  MCMC chains. For each parameterization and number of factors, the  ({\tt nChains}) heated chains are processed in parallel while swaps between pairs of chains are proposed. Parallelization is possible in the parameterization level as well, using the argument {\tt parallelModels}. This means that {\tt parallelModels} are running in parallel where each one of them runs {\tt nChains} chains in parallel, provided that the number of available threads is at least equal to $\mbox{{\tt nChains}}\times\mbox{{\tt parallelModels}}$. In order to parallelize our code, the {\tt doParallel} \citep{doparallel}, {\tt foreach} \citep{foreach} and {\tt doRNG} \citep{dorng} packages are imported.

The prior parameters $g, h, \alpha, \beta$ in Equations \eqref{eq:sigma_prior} and \eqref{eq:omega_prior} correspond to {\tt g}, {\tt h},{\tt alpha\_sigma} and {\tt beta\_sigma}, respectively, with a (common) default value equal to $0.5$. It is suggested to run the algorithm using {\tt normalize = TRUE}, in order to standardize the data before running the MCMC sampler. The default behaviour of our method is to normalize the data, thus, all reported estimates refer to the standardized dataset. In the case that the most probable number of mixture components is larger than 1, the ECR algorithm is applied in order to undo the label switching problem. Otherwise, the output is post-processed so that the generated parameters of the (single) alive component are switched to the first component of the overfitting mixture. 

The sampler will first run for {\tt warm\_up} iterations before starting to propose swaps between pairs of chain. By default, this stage consists of 5000 iterations. After that, each chain will run for a series of {\tt mCycles} MCMC cycles, each one consisting of {\tt nIterPerCycle} MCMC iterations (steps A, B, $\ldots$, G of the pseudocode). The updates of factors loadings according to \eqref{eq:lambda_fc} and \eqref{eq:lambda_fc2} at step B of the pseudocode are implemented using object-oriented programming using the {\tt Rcpp} and {\tt RcppArmadillo} libraries \citep{rcpp, rcpparmadillo}. At the end of each cycle, a swap between a pair of chains is proposed. 

Obviously, the total number of MCMC iterations is equal to $\mbox{{\tt warm\_up}}+\mbox{{\tt mCycles}}\times\mbox{{\tt nIterPerCycle}}$ and the first $\mbox{{\tt warm\_up}}+\mbox{{\tt burnCycles}}\times\mbox{{\tt nIterPerCycle}}$ iterations are discarded as burn-in. Given the default values of {\tt nIterPerCycle}, {\tt warm\_up} and {\tt overfittingInitialization}, choices between $50\leqslant \mbox{{\tt burnCycles}}\leqslant 500 <\mbox{{\tt mCycles}}\leqslant 1500$ are typical in our implementation (see also the convergence analysis in \cite{PAPASTAMOULIS2018220}).

While the function runs, some basic information is printed either on the screen (if {\tt parallelModels} is not enabled) or in separate text files inside the output folder (in the opposite case), such as the progress of the sampler as well as the acceptance rate of proposed swaps between chains. The output which is returned to the user is detailed in Table \ref{tab:output}. The full MCMC output of the selected model is returned as a list (named as {\tt mcmc}) consisting of {\tt mcmc} objects, a class imported from the {\tt coda} package \citep{coda}. In particular, {\tt mcmc} consists of the following:
\begin{itemize}
\item[$\bullet$] {\tt y}: object of class {\tt mcmc} containing the simulated factors.
\item[$\bullet$] {\tt w}: object of class {\tt mcmc} containing the simulated mixing proportions of the alive components, reordered according to ECR algorithm.
\item[$\bullet$] {\tt Lambda}: list containing objects of class {\tt mcmc} with the simulated factor loadings of the alive components, reordered according to ECR algorithm. Note that this particular parameter is not identifiable due to sign-switching across the MCMC trace. 
\item[$\bullet$] {\tt mu}: list containing objects of class {\tt mcmc} with the simulated marginal means of the alive components, reordered according to ECR algorithm.
\item[$\bullet$] {\tt z}: matrix of the simulated latent allocation variables of the mixture model, reordered according to ECR algorithm.
\item[$\bullet$] {\tt Sigma}: list containing objects of class {\tt mcmc} with the simulated variance of errors of the alive components, reordered according to ECR algorithm.
\item[$\bullet$] {\tt K\_all\_chains}: matrix of the simulated values of the number of alive components per chain.
\end{itemize}

The user can call the {\tt print}, {\tt summary} and {\tt plot} methods of the package in order to easily retrieve and visualize various summaries of the output, as exemplified in the next section.

\section{Examples}
This section illustrates our method. At first we demonstrate a typical implementation on two single simulated datasets and explain in detail the workflow. Then we perform an extensive simulation study for assesing the ability of the proposed method to recover the correct clustering and compare our findings to the {\tt pgmm} package \citep{McNicholas2008, doi:10.1093/bioinformatics/btq498, mcnicholas2010serial, pgmm}. Application to four publicly available datasets is provided next.

\subsection{Simulation study}\label{sec:sim}

\begin{figure*}[t]
\centering
\begin{tabular}{cc}
\includegraphics[scale=0.3]{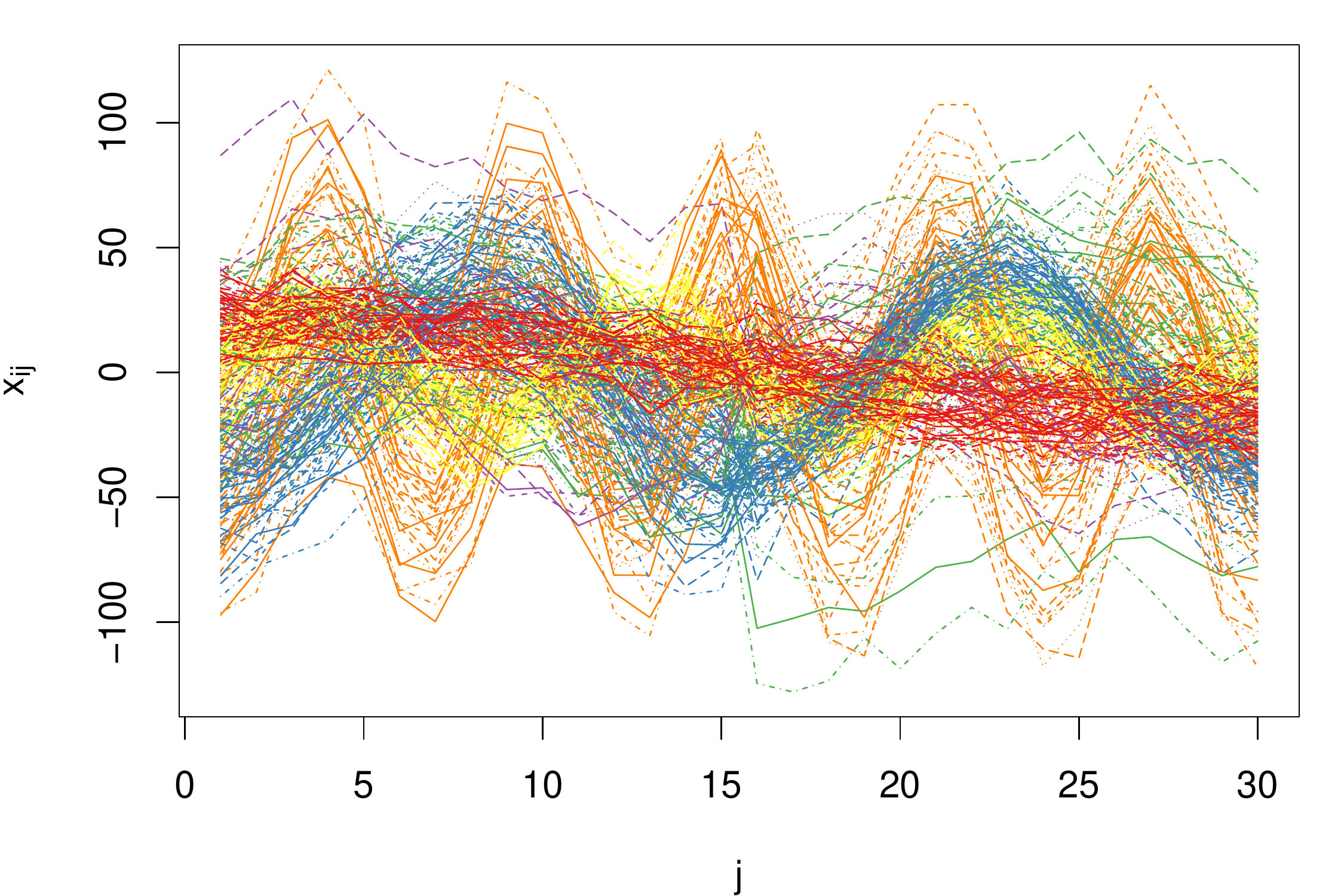} &
\includegraphics[scale=0.3]{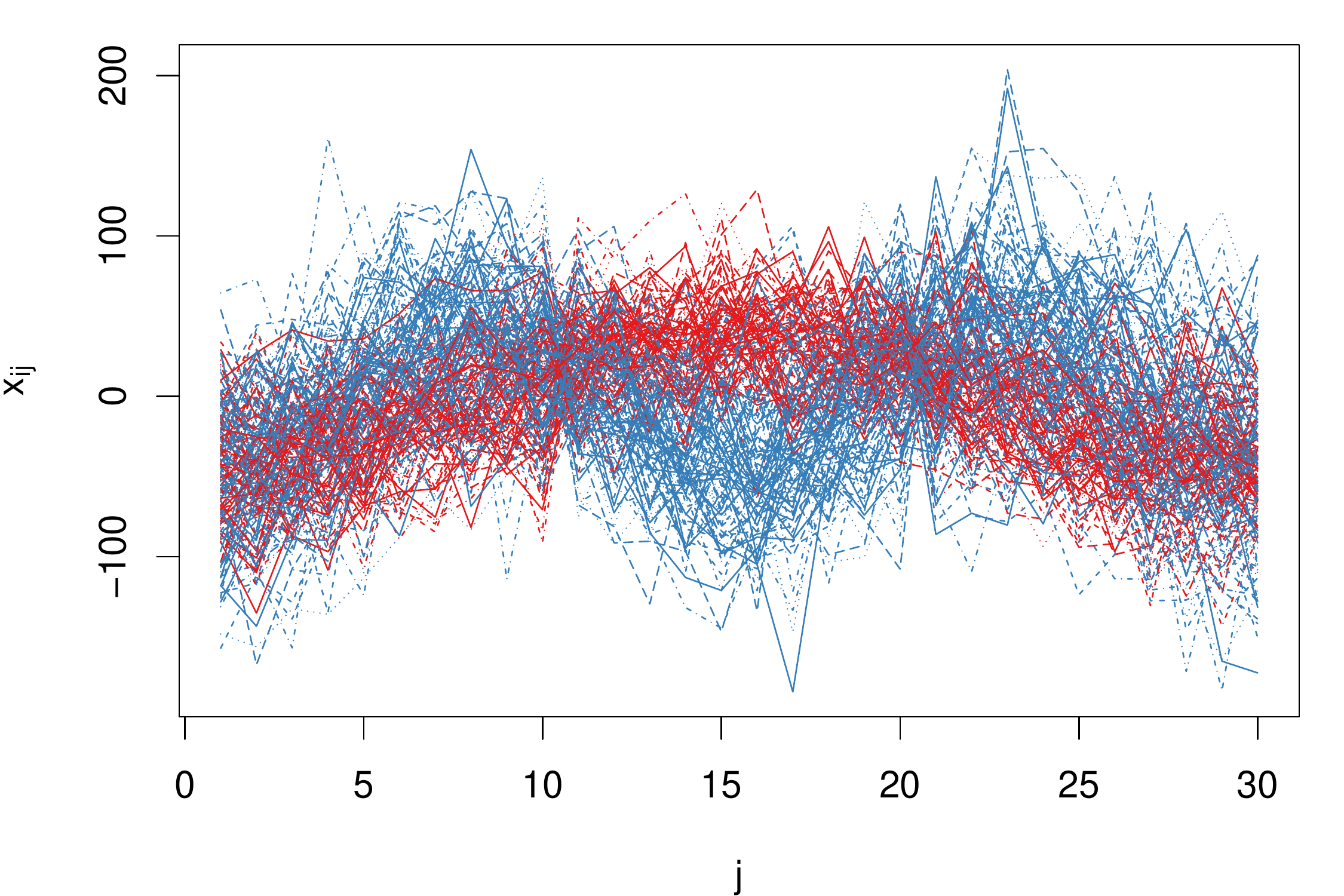}\\
(a) Dataset 1 & (b) Dataset 2
\end{tabular}
\caption{Simulated datasets of $p = 30$ variables consisting of $n = 300$ observations and $K = 6$ clusters (dataset 1) and $n = 200$, $K = 2$ (dataset 2). The colors display the ground truth classification of the data.}
\label{fig:sim_datasets}
\end{figure*}

We simulated synthetic data of $p = 30$ variables consisting of $n = 300$ observations and $K = 6$ clusters (dataset 1) and $n = 200$, $K = 2$ (dataset 2), as shown in Figure \ref{fig:sim_datasets}. Both of them were generated using MFA models with $q = 2$ (dataset 1) and $q = 3$ (dataset 2) factors. The two datasets exhibit different characteristics: the variance of errors per cluster ($\bs\Sigma_k$) is significantly larger in dataset 2 compared to dataset 1. In addition, the selection of factor loadings in dataset 2 result to more complex covariance structure. The generating mechanism, described in detail in \cite{PAPASTAMOULIS2018220}, is available in the {\tt fabMix} package via the {\tt simData()} and {\tt simData2()} functions, as shown below.
\begin{verbatim}
> library('fabMix')
#  dataset 1
> set.seed(1)
> n = sample(100*(1:10), 1) # sample size
> q = sample(1:3, 1)  # number of factors
> K = sample(1:10, 1) # number of clusters
# results to n = 300, q = 2, K = 6
> p = 30    # number of variables
# inverse variance of errors
> sINV <- array(data = NA, dim = c(K,p))  
> for(k in 1:K){sINV[k,] <- 1/(1+20*log(k+1))}
> dataset1 <- simData(sameSigma=FALSE, K.true=K, 
+ n = n, q = q, p = p, sINV_values=sINV)
# synthetic dataset 2
> set.seed(30)
> n = 200; q = 3; K = 2; p = 30
# inverse variance of errors
> sINV <- array(data = NA, dim = c(K,p))   
> for(k in 1:K){sINV[k,] <- 1/(1+1000*log(k+1))}
> dataset2 <- simData2(sameSigma=FALSE, K.true=K, 
+ n = n, q = q, p = p, sINV_values = sINV)
\end{verbatim} 
Next we estimate the 8 overfitting Bayesian MFA models with $K_{\mbox{max}}=20$ mixture components assuming that the number of factors ranges in the set $1\leqslant q\leqslant 5$. The MCMC sampler runs {\tt nChains = 4} heated chains, each one consisting of {\tt mCycles = 700} cycles, while the first {\tt burnCycles = 100} are discarded. Recall that each MCMC cycle consists of {\tt nIterPerCycle = 10} usual MCMC iterations and that there is an additional warm up period of the MCMC sampler (before starting to propose chain swaps) corresponding to 5000 usual MCMC iterations. 

\begin{verbatim}
> Kmax <- 20    # number of components
> nChains <- 4  # number of chains
> qRange <- 1:5 # number of factors
#       Run fabMix() for dataset 1
set.seed(1)
> fm1 <- fabMix(nChains = nChains, 
+ rawData = dataset1$data, outDir = "tmp1", 
+ Kmax = Kmax, mCycles = 700, burnCycles = 100, 
+ q = qRange, parallelModels = 4) 
#       Run fabMix() for dataset 2
set.seed(1)
> fm2 <- fabMix(nChains = nChains, 
+ rawData = dataset2$data, outDir = "tmp2", 
+ Kmax = Kmax, mCycles = 700, burnCycles = 100, 
+ q = qRange, parallelModels = 4) 
\end{verbatim}
The argument {\tt parallelModels = 4} implies that 4 parameterizations will be processed in parallel. In addition, each model will use {\tt nChains = 4} threads to run in parallel the specified number of chains. Our jobscript used 16 threads so in this case the $\mbox{{\tt parallelModels}}\times{{\tt nChains}} = 16$ jobs are efficiently allocated. 

\subsubsection{Methods for printing, summarizing and plotting the output}

The {\tt print} method for a {\tt fabMix.object} displays some basic information for a given run of the {\tt fabMix} function. The following output corresponds to the first dataset. 

\begin{verbatim}
> print(fm1)
* Run information: 
      Number of fitted models: 
      (5 factor levels) x (8 parameterizations)
       = 40 models.
      Selected model: UUC model with K = 6 
      clusters and q = 2 factors.

* Maximum A Posteriori (MAP) number of `alive' 
clusters and selected number of factors (BIC) 
per model: 
  model K_MAP K_MAP_prob q  BIC_q chain_swap
1   UUU     4       1.00 3 4295.0      8.43%
2   CUU     5       0.70 3 4248.9     13.29%
3   UCU     7       0.71 2 3345.8     25.86%
4   CCU    13       0.70 2 3410.5     93.71%
5   UCC     7       0.55 2 3270.7     22.71%
6   UUC     6       1.00 2 2274.5     19.43%
7   CUC    10       0.41 3 2868.3     78.43%
8   CCC    13       0.52 2 3378.3     92.43%

* Estimated number of observations per cluster
 (selected model): 
label
 4  7 13 14 15 17 
60 55 41 72 50 22 
\end{verbatim}

\noindent
The following output corresponds to the {\tt print} method for the {\tt fabMix} function for the second dataset.
\begin{verbatim}
> print(fm2)
* Run information: 
      Number of fitted models:
      (5 factor levels) x (8 parameterizations)
       = 40 models.
      Selected model: UUC model with K = 2 
      clusters and q = 2 factors.

* Maximum A Posteriori (MAP) number of `alive' 
clusters and selected number of factors (BIC)
per model: 
  model K_MAP K_MAP_prob q   BIC_q chain_swaps
1   UUU     2       1.00 2 14776.6       4.86%
2   CUU     2       1.00 2 14676.0          3%
3   UCU     2       0.64 3 15043.1       4.57%
4   CCU     3       0.48 3 14836.2      12.71%
5   UCC     3       0.64 2 15141.9       4.29%
6   UUC     2       0.95 2 14558.5       3.14%
7   CUC     3       0.85 3 14598.7       4.14%
8   CCC     4       0.44 3 14851.6      11.14%

* Estimated number of observations per cluster
 (selected model): 
label
  6  20 
113  87 
\end{verbatim}

We conclude that the selected models correspond to the UUC parameterization with $K = 6$ clusters and $q = 2$ factors for dataset 1 and $K = 2$, $q = 2$ for dataset 2. The selected number of clusters and factors for the whole range of 8 models is displayed next, along with the estimated posterior probability of the number of alive clusters per model ({\tt K\_MAP\_prob}), the value of the BIC for the selected number of factors ({\tt BIC\_q}) as well as the proportion of the accepted swaps between the heated MCMC chains in the last column. The frequency table of the estimated single best clustering of the datasets is displayed in the last field. We note that the labels of the frequency table correspond to the labels of the alive components of the overfitting mixture model, that is, components 4, 7, 13, 14, 15, and 17 for dataset 1 and components 6 and 20 for dataset 2. Clearly, these labels can be renamed to $1,2,3,4,5,6$ and $1,2$ respectively, but we prefer to retain the raw output of the sampler as a reminder of the fact that it corresponds to the alive components of the overfitted mixture model. 

The {\tt summary} method of the {\tt fabMix} package summarizes the MCMC output for the selected model by calculating posterior means and quantiles for the mixing proportions, marginal means and the covariance matrix per (alive) cluster. A snippet of the output for dataset 2 is shown below. 

\begin{verbatim}
> s <- summary(fm2)
* `Alive' cluster labels: 
[1] "6"  "20"

* Posterior mean of the mixing proportions: 
   6   20 
0.58 0.42 

* Posterior mean of the marginal means: 
        Cluster label
Variable     6    20
     V1  -0.06  0.08
     V2  -0.02  0.02
     ...............
     V30 -0.01  0.00

* Posterior mean of the covariance matrix: 

   Covariance matrix for cluster `6': 
       V1    V2  ...   V30
V1   1.12  0.53  ... -0.18
V2   0.53  1.11  ... -0.14
..........................
V30 -0.18 -0.14  ...  1.27

   Covariance matrix for cluster `20': 
       V1    V2  ...   V30
V1   0.66  0.39  ... -0.05
V2   0.39  0.88  ... -0.03
..........................
V30 -0.05 -0.03  ...  0.57

Quantiles for each parameter: 
                quantile
parameter        2.5%   25%   50%  75%  97.5%
  weight_6       0.51  0.55  0.58  0.51  0.65
  weight_20      0.35  0.40  0.42  0.45  0.50
  mean_6_V1     -0.27 -0.13 -0.06  0.01  0.13
  mean_20_V1    -0.09  0.03  0.08  0.14  0.25
  mean_6_V2     -0.22 -0.09 -0.02  0.04  0.15
  mean_20_V2    -0.17 -0.05  0.02  0.08  0.19
  ...........................................
  mean_6_V30    -0.21 -0.08 -0.01  0.07  0.20
  mean_20_V30   -0.16 -0.06  0.00  0.06  0.17
  cov_6_V1_V1    0.89  1.03  1.10  1.19  1.46
  cov_20_V1_V1   0.50  0.59  0.64  0.72  0.85
  cov_6_V1_V2    0.37  0.46  0.52  0.59  0.78
  cov_20_V1_V2   0.24  0.32  0.38  0.44  0.56
  ...........................................
  cov_6_V1_V30  -0.37 -0.24 -0.18 -0.12  0.01
  cov_20_V1_V30 -0.16 -0.08 -0.05 -0.01  0.06
  ...........................................
  cov_6_V2_V30  -0.34 -0.21 -0.13 -0.08  0.04
  cov_20_V2_V30 -0.17 -0.07 -0.03  0.03  0.12
  ...........................................
  cov_6_V30_V30  1.03  1.18  1.26  1.37  1.62
  cov_20_V30_V30 0.45  0.51  0.56  0.61  0.73
\end{verbatim}
\noindent
The printed output is also returned to the user via {\tt s\$posterior\_means} and {\tt s\$quantiles}.

\begin{figure}[t]
\centering
\begin{tabular}{c}
\includegraphics[scale=0.35]{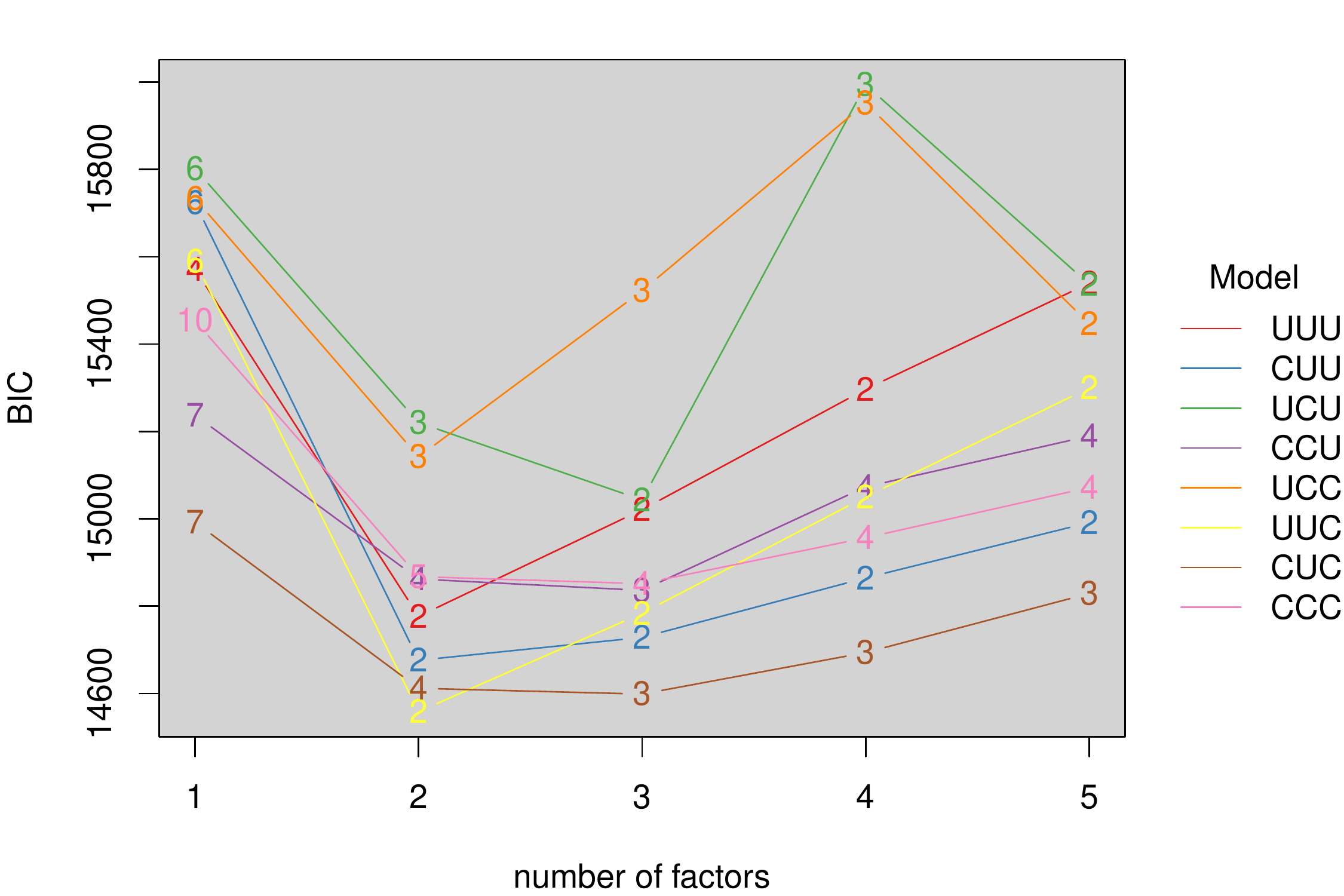} \\
(a) Dataset 1 \\
\includegraphics[scale=0.35]{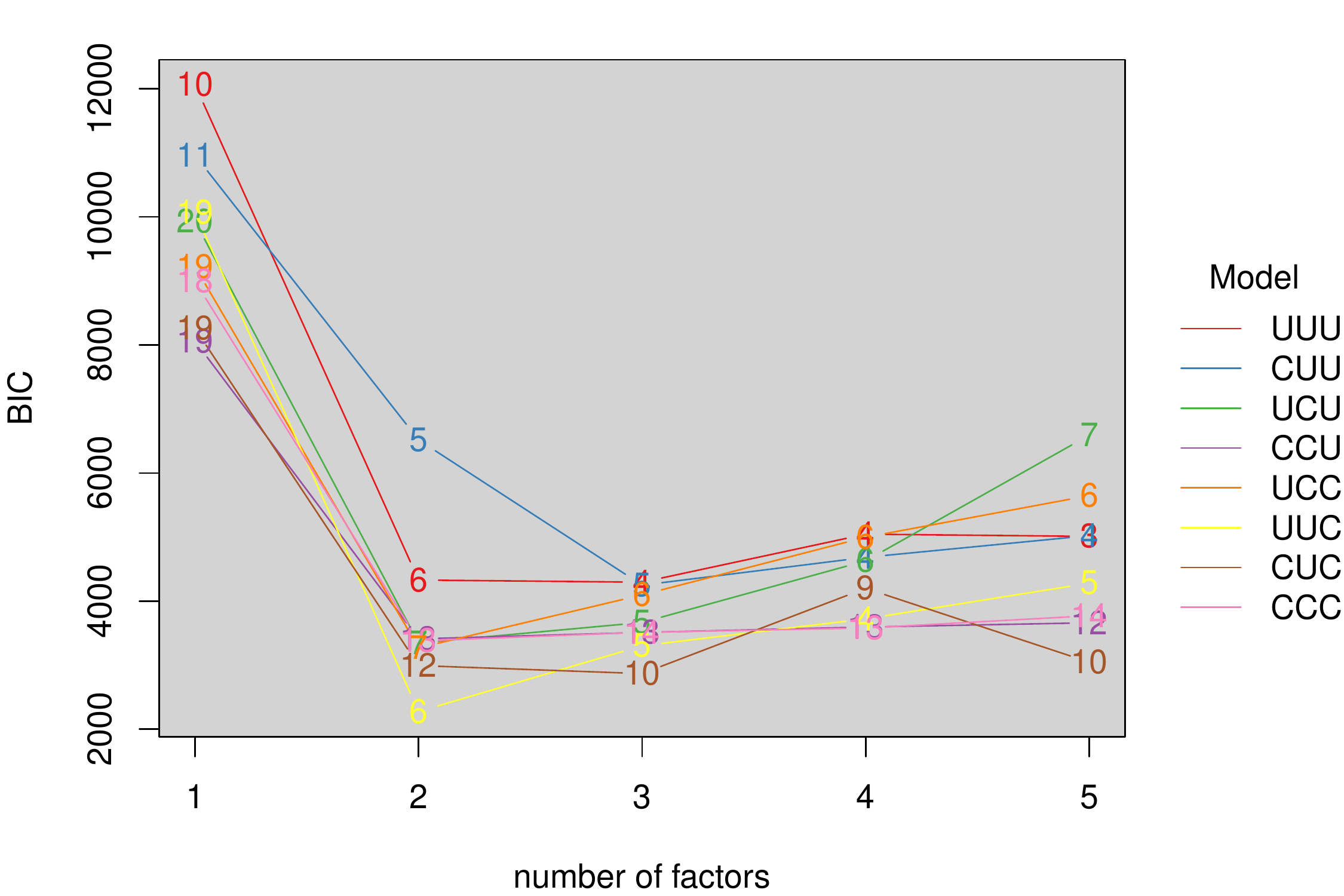}\\
 (b) Dataset 2
\end{tabular}
\caption{BIC values per parameterization and factor level using the {\tt plot(fabMix.object)} method.}
\label{fig:bic}
\end{figure}

The {\tt plot()} method of the package generates the following types of graphics output: 
\begin{enumerate}
\item[(1)] Plot of the BIC values per factor level and parameterization.
\item[(2)] Plot of the posterior means of marginal means ($\bs\mu_k$) per (alive) cluster and Highest Density Intervals of the corresponding normal distribution along with its assigned data. 
\item[(3)] The coordinate projection plot of the {\tt mclust} package \citep{mclust1, mclust2}, that is, a scatterplot of the assigned data per cluster for each pair of variables.
\item[(4)] Visualization of the posterior mean of the correlation matrix per cluster using the {\tt corrplot} package.
\item[(5)] The MAP estimate of the factor loadings ($\bs\Lambda_k$) per (alive) cluster.
\end{enumerate}

The following commands produce plot (1) for datasets 1 and 2.
\begin{verbatim}
> plot(fm1, what = 'BIC')
> plot(fm2, what = 'BIC')
\end{verbatim}
The produced plots are shown in Figure \ref{fig:bic}. Note that each point in the plot is labeled by an integer, which corresponds to the MAP number of alive components for the specific combination of factors and parameterization.

\begin{figure}[ht]
\centering
\begin{tabular}{cc}
\hspace{-5ex}\includegraphics[scale=0.3]{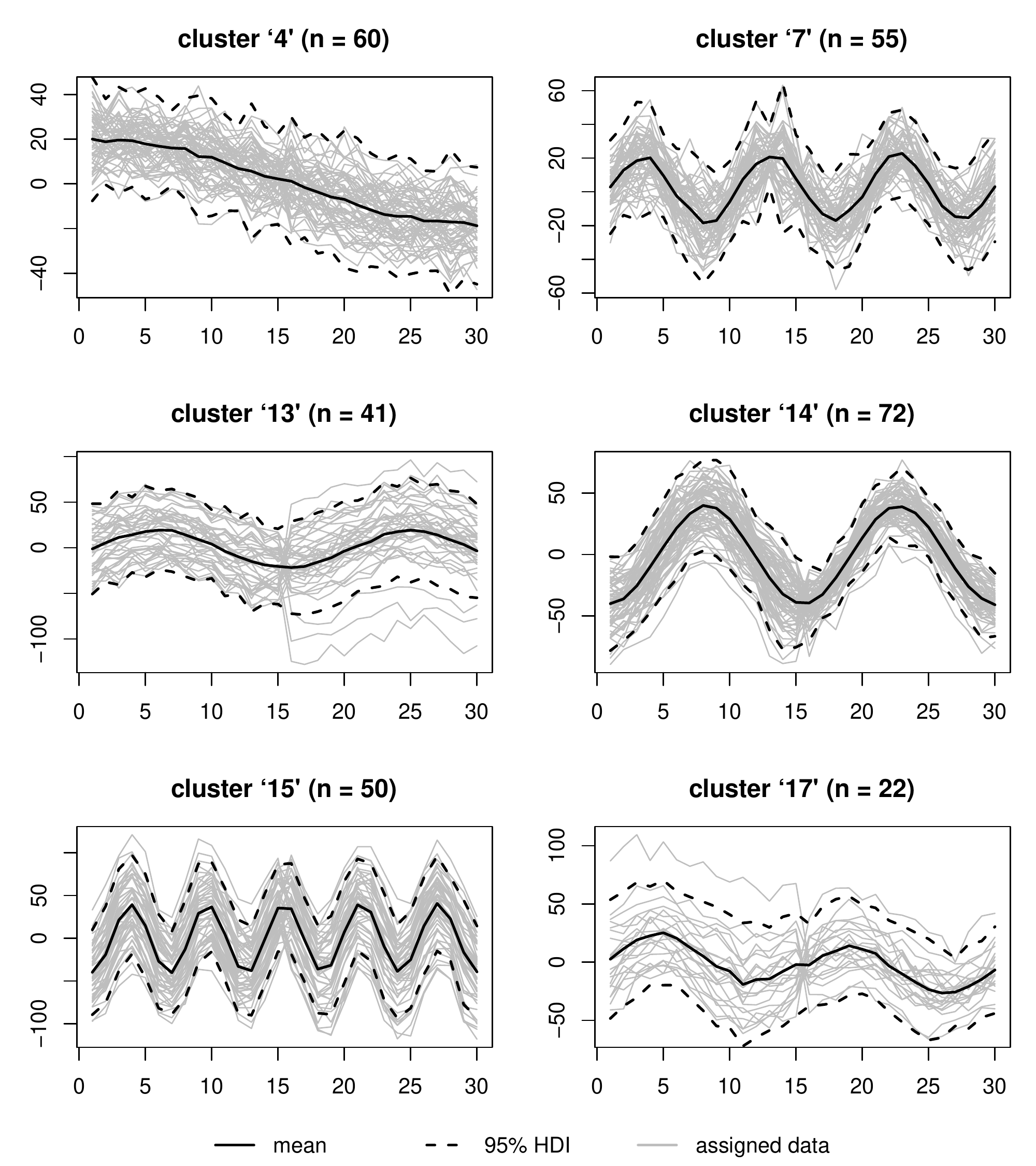} &
\hspace{-5ex}\includegraphics[scale=0.3]{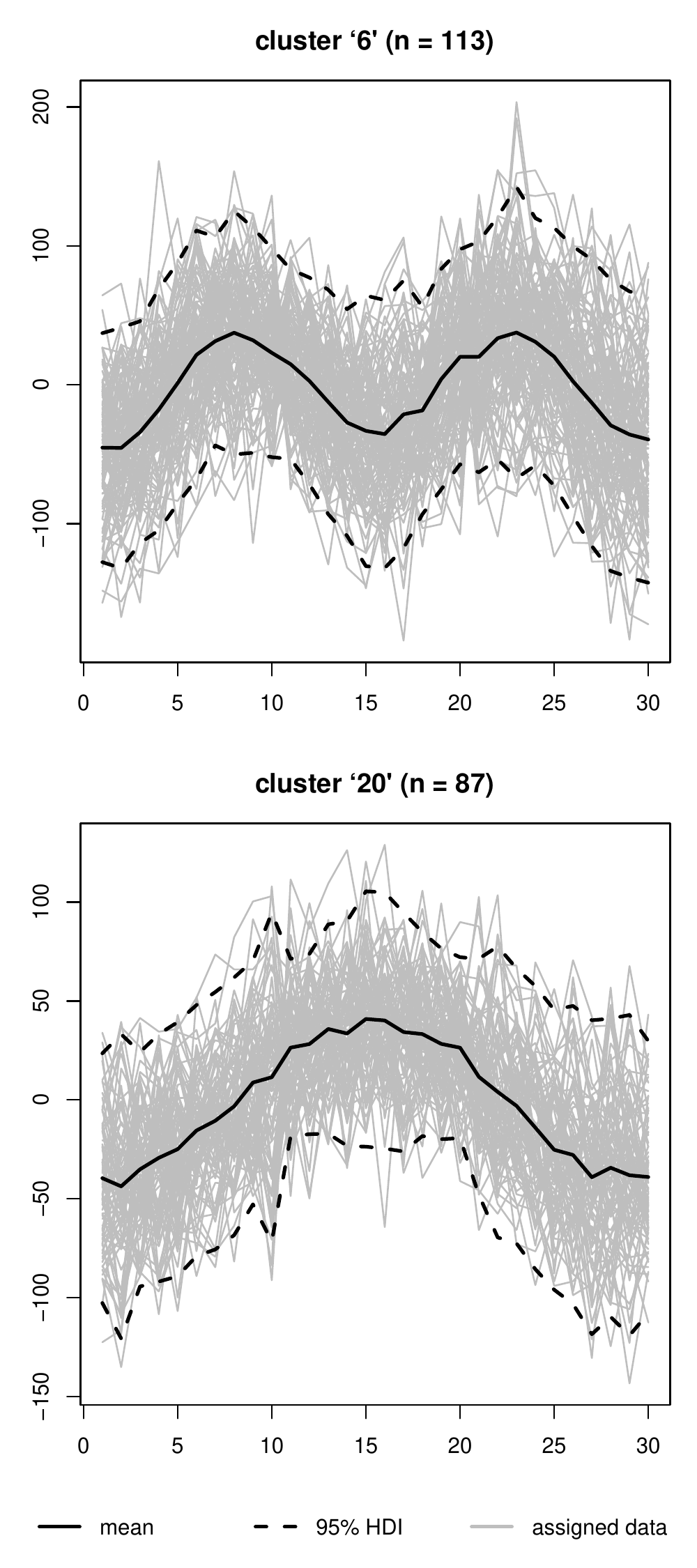}\\
(a) Dataset 1 & (b) Dataset 2
\end{tabular}
\caption{Marginal mean with $95\%$ Highest Density Interval and the corresponding assigned data per alive cluster using the {\tt plot(fabMix.object)} method.}
\label{fig:classification}
\end{figure}

The following commands produce plot (2) for datasets 1 and 2.
\begin{verbatim}
> plot(fm1, what = 'classification_matplot', 
+ class_mfrow = c(3,2), confidence = 0.95)
> plot(fm2, what = 'classification_matplot', 
+ class_mfrow = c(2,1), confidence = 0.95)
\end{verbatim}
The created plots are shown in Figure \ref{fig:classification}. The {\tt class\_mfrow} arguments control the rows and columns of the layout and it should consists of 2 integers with their product equal to the selected number of (alive) clusters. In addition, a legend is placed on the bottom of the layout. The value(s) in the {\tt confidence} argument draws the Highest Density Interval(s) of the estimated normal distribution. Note that these plots display the original and not the scaled dataset which is used in the MCMC sampler. Therefore, the central curve and confidence limits displayed in the specific plot correspond to the mean and variance (multiplied by the appropriate quantile of the standard Normal distribution) of the random variables arising by applying the inverse of the z-transformation on the MCMC estimates reported by the {\tt fabMix} function.

\begin{figure}[h]
\centering
\begin{tabular}{cc}
\hspace{-5ex}\includegraphics[scale=0.28, page = 1]{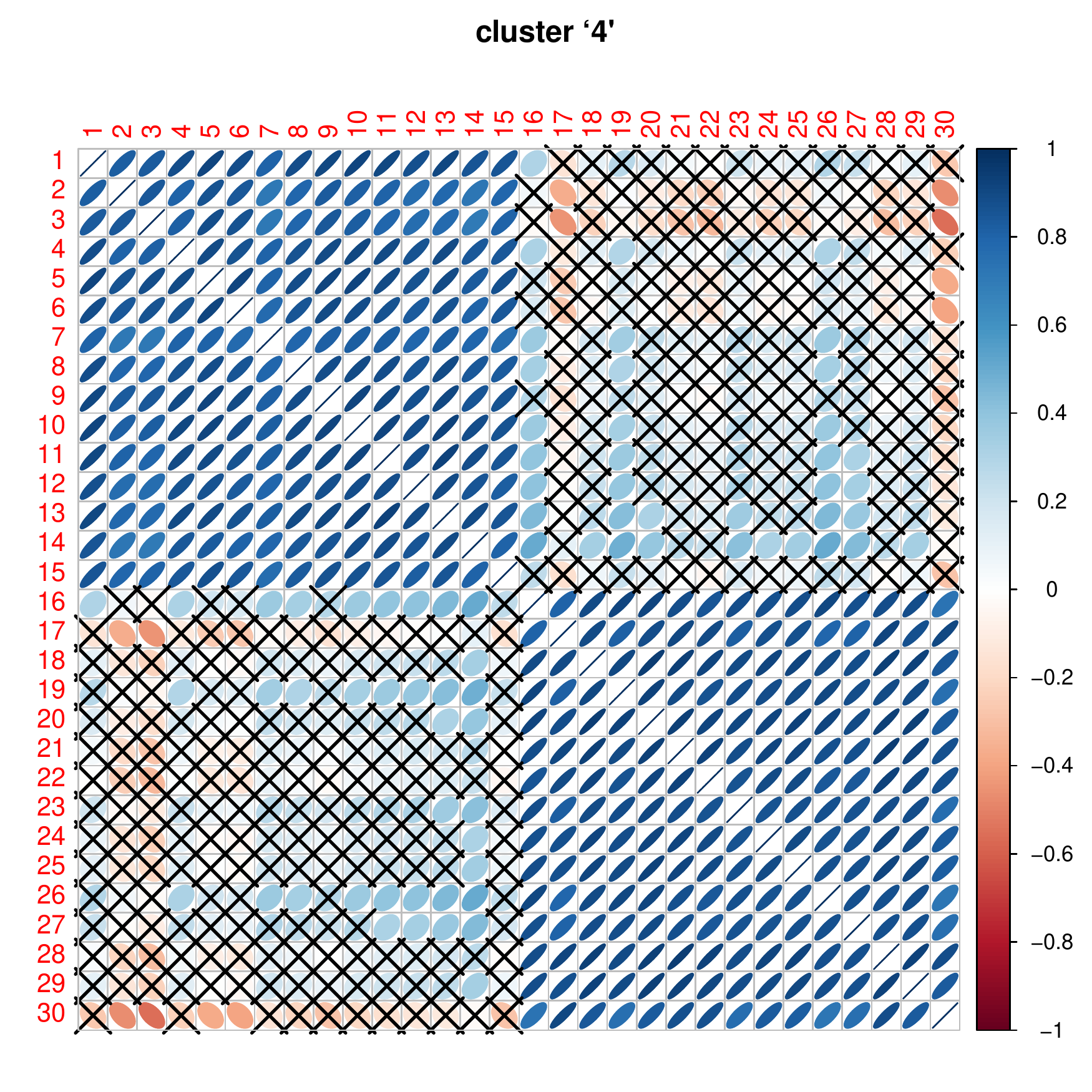} &
\hspace{-7.5ex}\includegraphics[scale=0.28, page = 1]{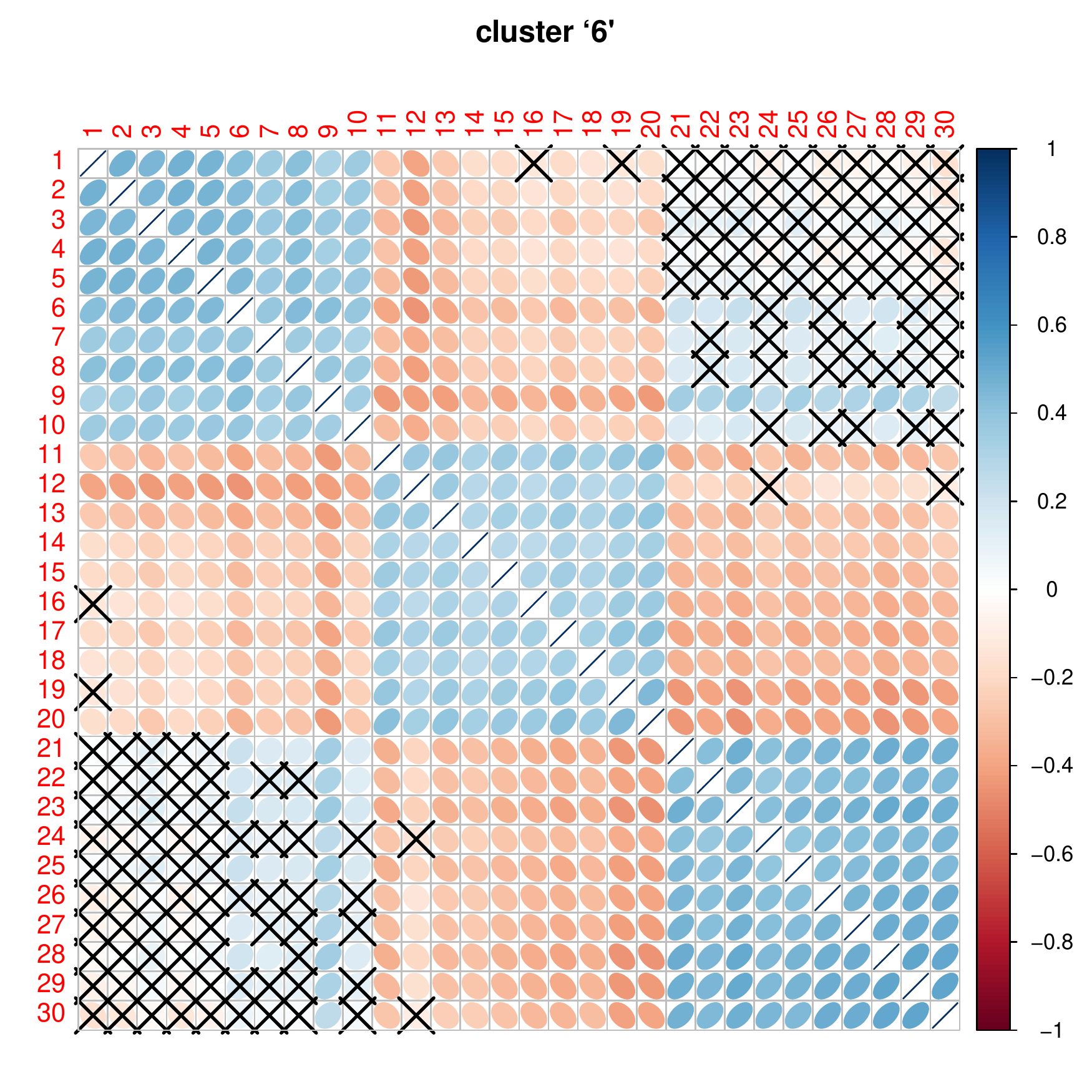}
\\
(a) Dataset 1 & (b) Dataset 2
\end{tabular}
\caption{Correlation matrix for the first (alive) cluster of each dataset.}
\label{fig:lambda}
\end{figure}

Figure \ref{fig:lambda} visualizes the correlation matrix for the first cluster of each dataset, using the {\tt corrplot} package. The argument $\mbox{{\tt sig\_correlation}} = \alpha$ is used for marking cases where the equally tailed $(1-\alpha)$ Bayesian credible interval contains zero. The following commands generate the plots in Figure \ref{fig:lambda}. 
\begin{verbatim}
> plot(fm1, what = 'correlation', 
+	sig_correlation = 0.05)
> plot(fm2, what = 'correlation', 
+ sig_correlation = 0.05)
\end{verbatim}

%
%

\begin{figure*}[p]
\centering
\begin{tabular}{cc}
\includegraphics[scale=0.35]{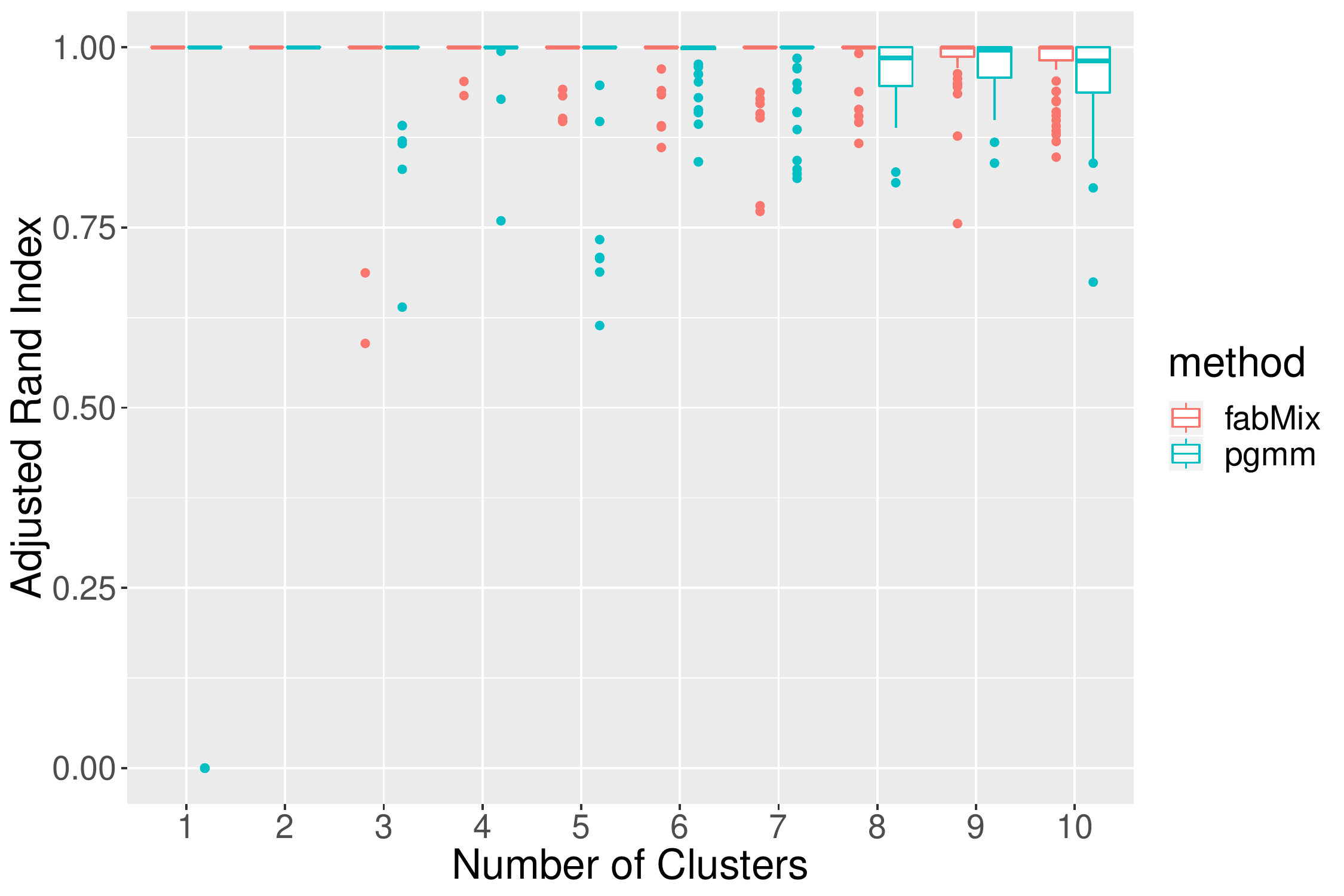} &\includegraphics[scale=0.35]{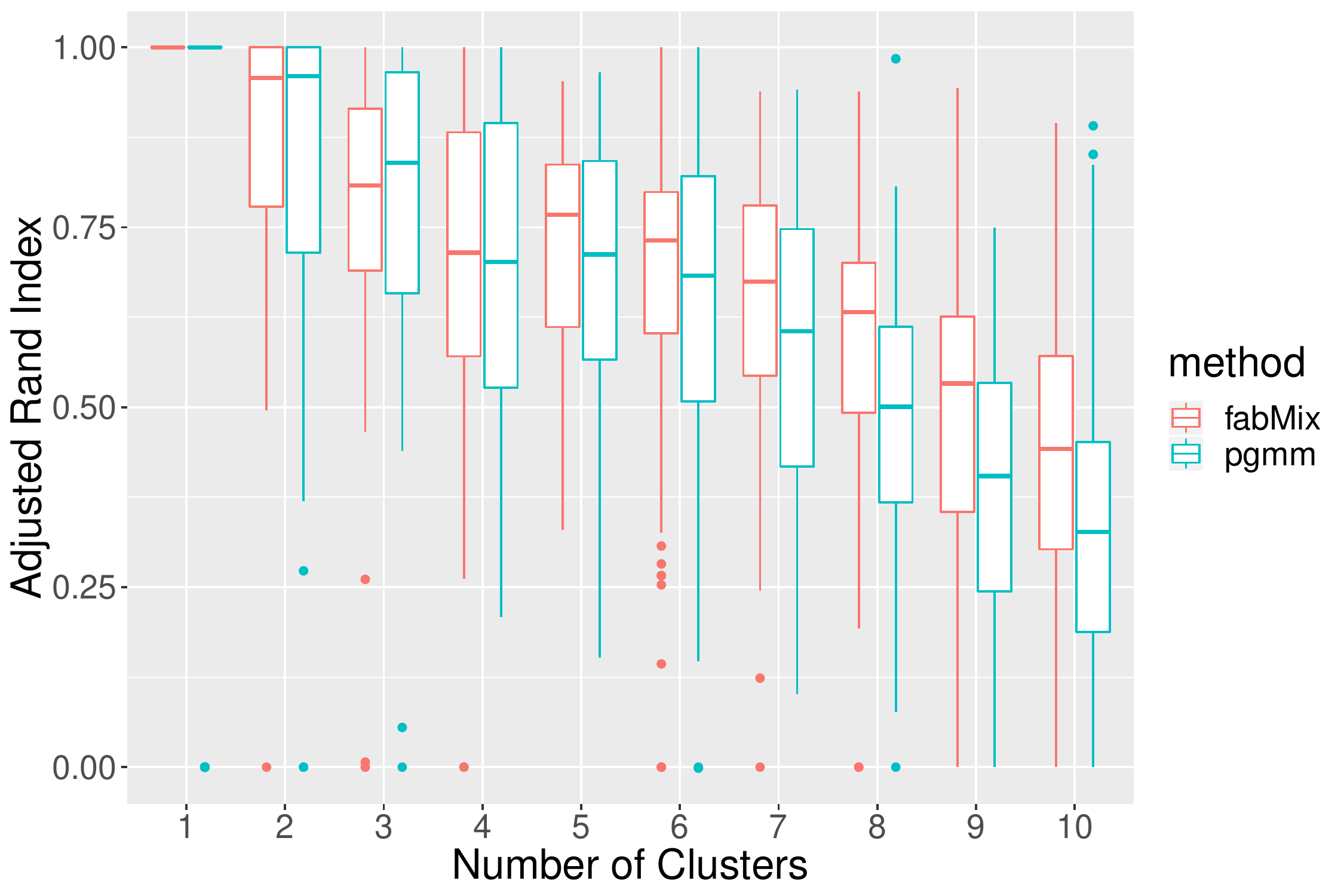}  \\
\includegraphics[scale=0.35]{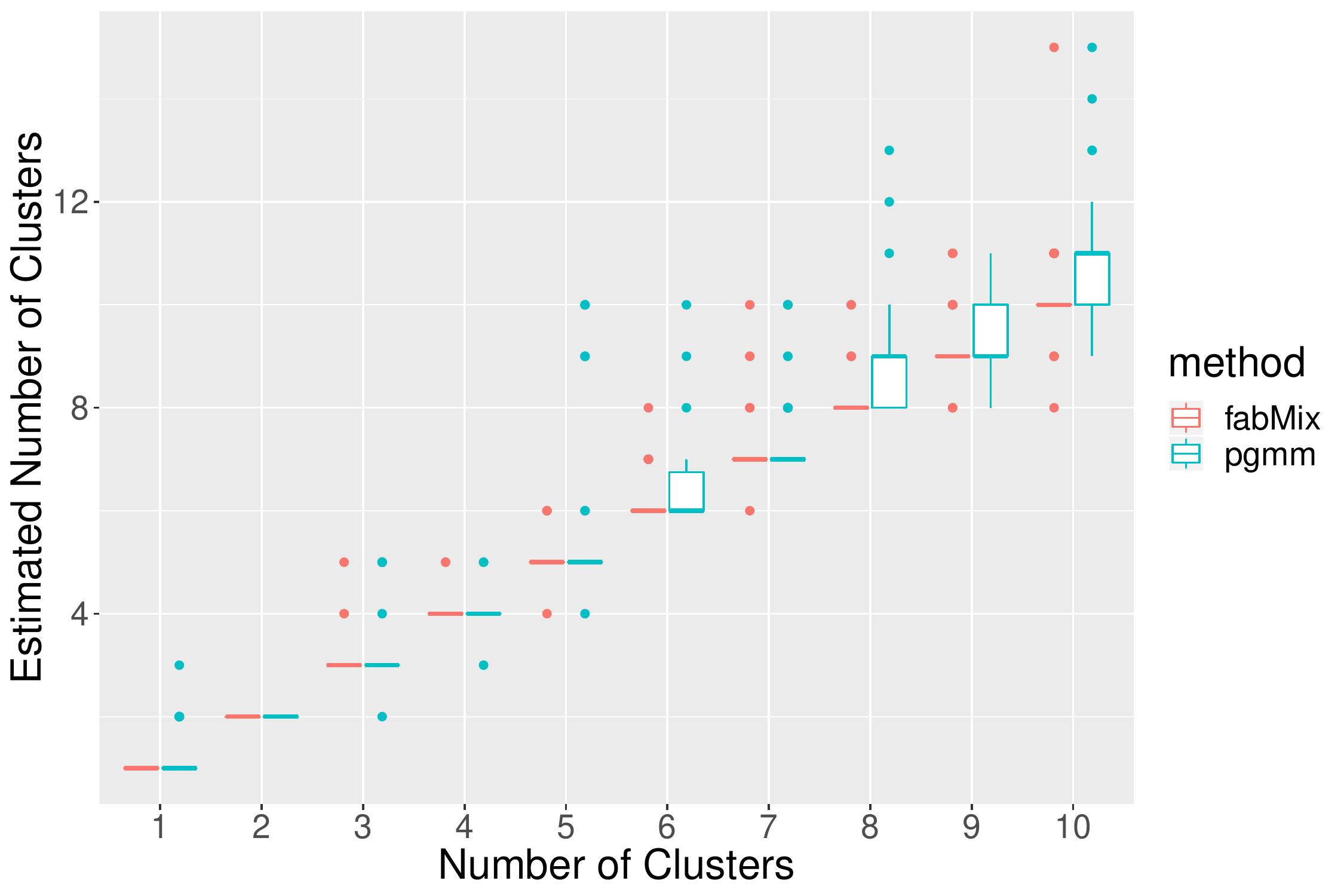} &\includegraphics[scale=0.35]{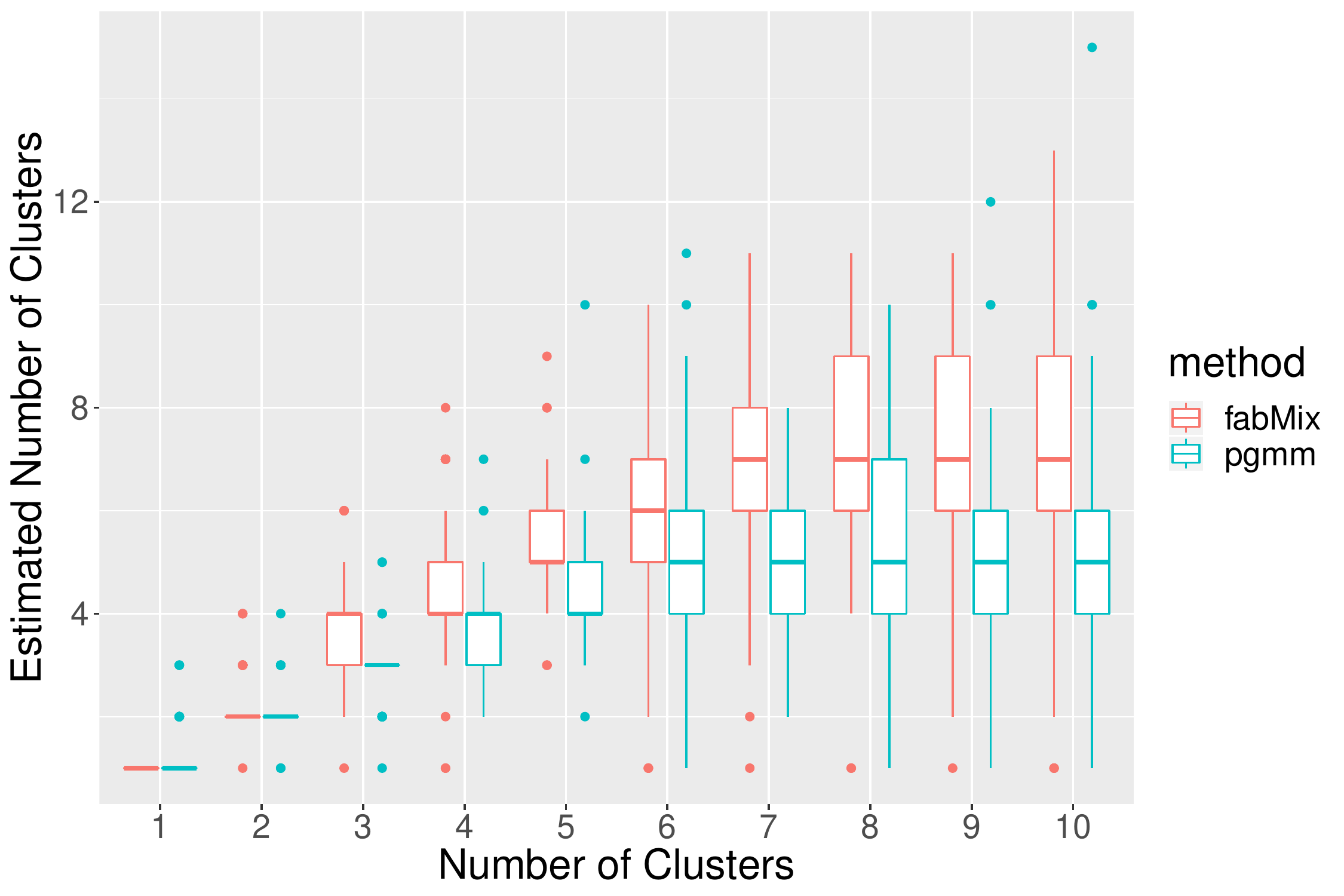}  \\
\includegraphics[scale=0.35]{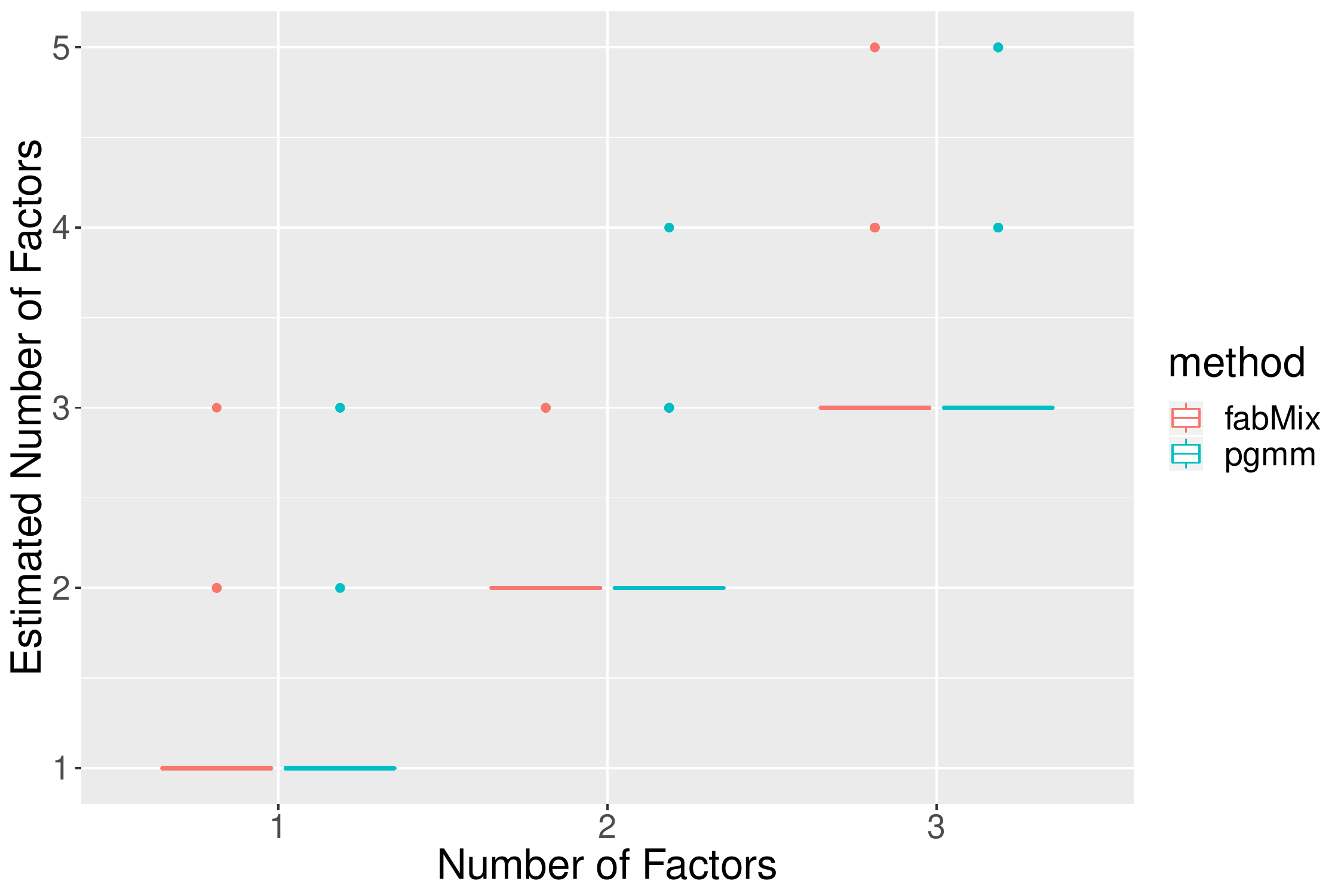} &\includegraphics[scale=0.35]{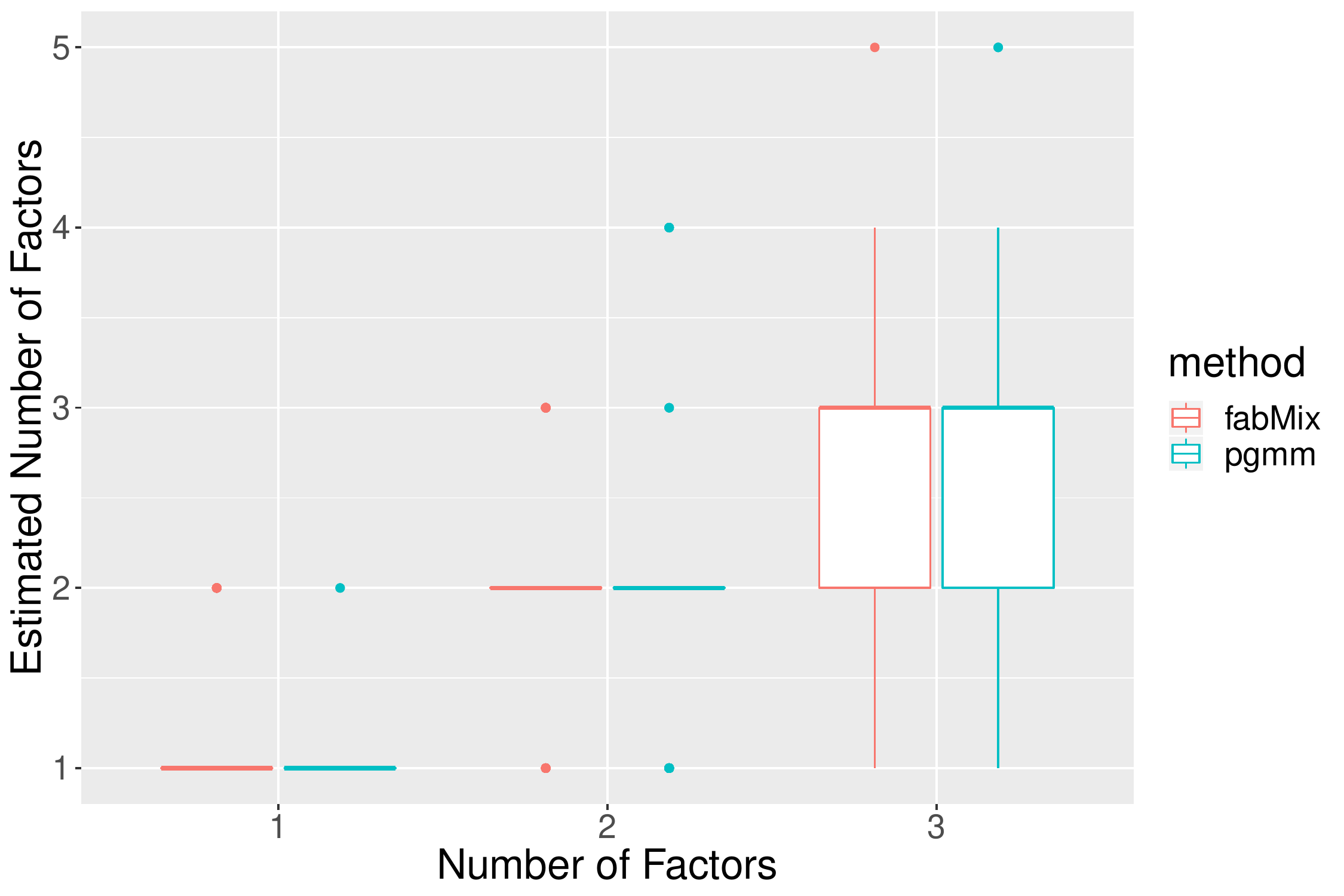}  \\
\includegraphics[scale=0.35]{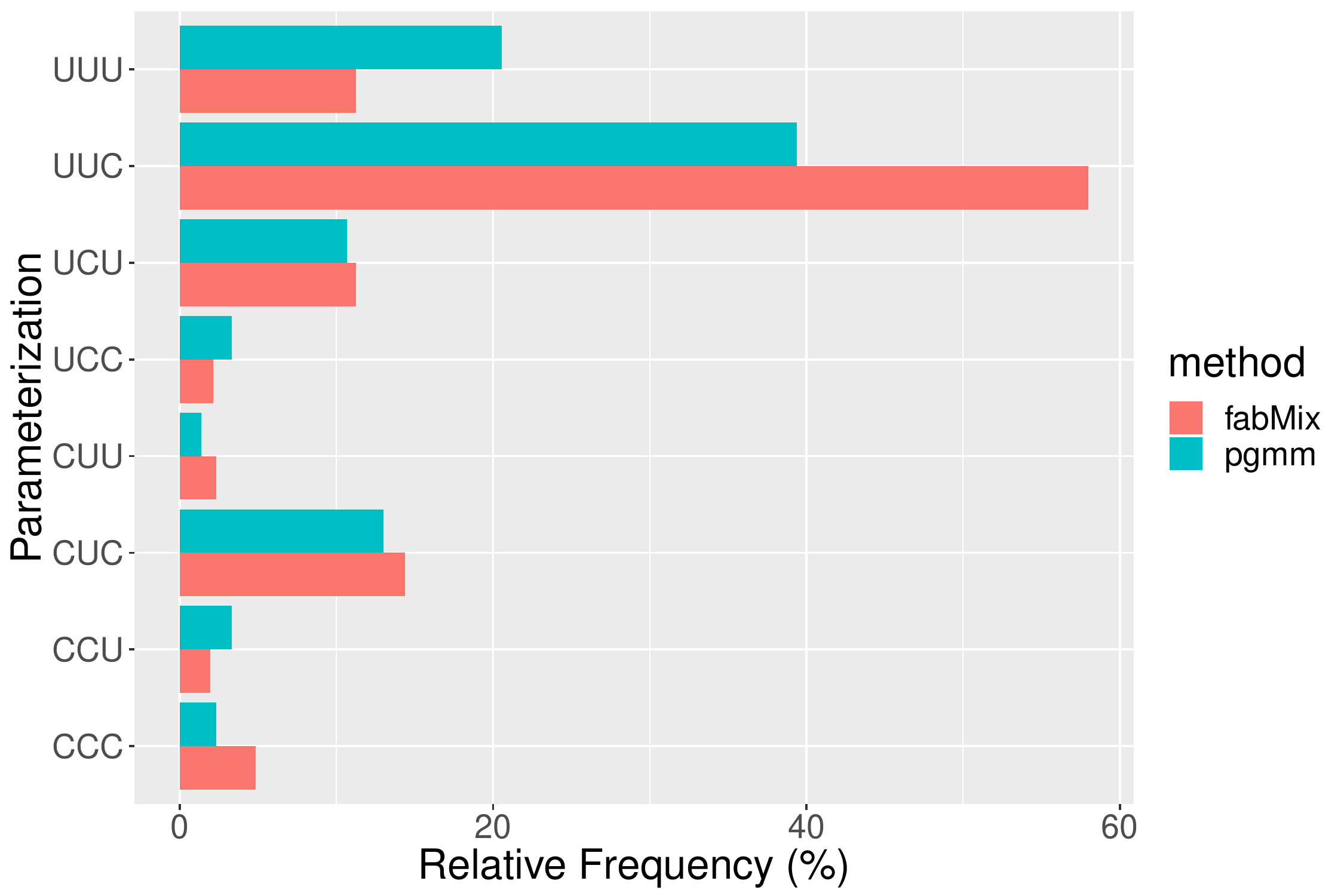} &  \includegraphics[scale=0.35]{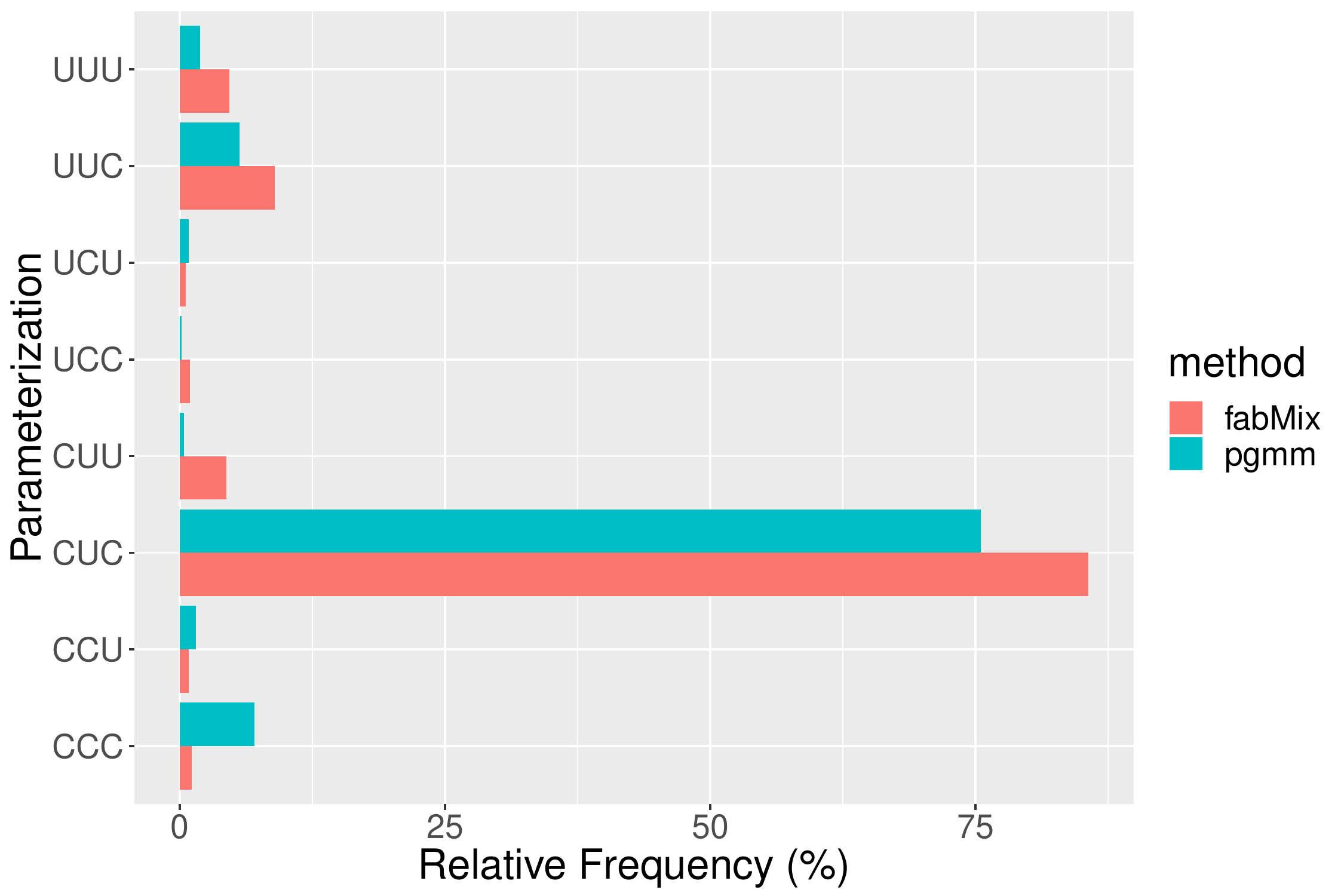}\\
(a) Scenario 1 & (b) Scenario 2
\end{tabular}
\caption{Adjusted Rand Index (first row), estimated number of clusters (second row), estimated number of factors (third row) and selected parameterization (last row) for various replications of Scenarios 1 and 2 with varying number of clusters and factors. In all cases the sample size is drawn randomly in the set $\{100, 200, \ldots,1000\}$.}
\label{fig:many}
\end{figure*}

\subsubsection{Assessing clustering accuracy and comparison with {\tt pgmm}}

In this section we compare our findings against the ground-truth in simulated datasets and also compare against the {\tt pgmm} package, considering the same range of clusters and factors per dataset. For each combination of number of factors, components and parameterization, the {\tt pgmmEM()} algorithm was initialized using 3 random starting values as well as the K-means clustering algorithm, that is, 4 different starts in total. Note that the number of different starts of the EM algorithm is set equal to number of parallel chains in the MCMC algorithm. The input data is standardized in both algorithms. 

As shown in Table \ref{tab:tab1}, the adjusted Rand index (ARI) \citep{doi:10.1080/01621459.1971.10482356} between {\tt fabMix} and the ground-truth classification is equal to 1 and 0.98 for simulated dataset 1 and 2, respectively. The corresponding ARI for {\tt pgmm} equals to 0.98 and 0.88, respectively. In both cases our method finds the correct number of clusters, however {\tt pgmm} overestimates $K$ in dataset 1. Both methods select the UUC parameterization in dataset 1, but in dataset 2 different models are selected (UUC by {\tt fabMix} and CUC by {\tt pgmm}).

\begin{table}[t]
\centering
\caption{Selected number of clusters, factors, parameterization and adjusted rand index for simulated data 1 and 2.}
\begin{tabular}{p{14mm}p{2mm}p{2mm}p{5mm}p{5mm}p{2mm}p{2mm}p{5mm}p{5mm}}
\toprule
\multirow{2}{*}{Data $(K,q)$} &
\multicolumn{4}{c}{{\tt fabMix}} &
\multicolumn{4}{c}{{\tt pgmm}}\\
 \cline{2-9}\\[-0.9em]
 & $\widehat{K}$ & $\widehat{q}$ & model & ARI &  $\widehat{K}$ & $\widehat{q}$ & model & ARI \\
\midrule
1 $(6,2)$ &  6 & 2 & UUC & 1 & 7 & 2 & UUC & .95 \\
2 $(2,3)$&  2 & 2 & UUC & .98 & 2 & 2 & CUC & .88\\
\bottomrule
\end{tabular}
\label{tab:tab1}
\end{table}

The selected number of factors equals 2, however in dataset 2 the ``true'' number of factors equals 3. The underestimation of the number of factors in dataset 2 remains true for a wide range of similar data: in particular we generated synthetic datasets with identical parameter values as the ones in dataset 2 but each time the sample size was increasing by 200 observations. We observed that the correct number of factors is returned when $n \geqslant 1600$ for {\tt fabMix} and $n \geqslant 1800$ for {\tt pgmm}.

Next we replicate the two distinct simulation procedures (according to the {\tt simData()} and {\tt simData2()} functions of the package) used to generate the previously described datasets, but considering that $1\leqslant K \leqslant 10$ (true number of clusters) and $1\leqslant q \leqslant 3$ (true number of factors). The number of variables remains the same as before, that is, $p=30$ and the sample size is drawn uniformly at random in the set $\{100,200,\ldots,1000\}$. We will use the terms 'Scenario 1' and 'Scenario 2' to label the two different simulation procedures. In Scenario 1 the diagonal of the variance of errors is generated as $\sigma_{kr}^{2} = 1 + 20\log(k+1)$, $r = 1,\ldots,p$, whereas in Scenario 2: $\sigma_{kr}^{2} = 1 + u_r\log(k+1)$, where $u_r\sim \mbox{Uniform}(500,1000)$, $r = 1,\ldots,p$; $k=1,\ldots,K$. In general, Scenario 1 generates datasets with well separated clusters. On the other hand, the amount of error variance in Scenario 2 makes the clusters less separated. For a given simulated dataset with $K_{\mbox{true}}$ clusters and $q_{\mbox{true}}$ factors, we are considering that the total number of components in the overfitting mixture model ({\tt fabMix}) as well as the maximum number of components fitted from {\tt pgmm} is set equal to $K_{\mbox{max}} = K_{\mbox{true}} + 6$ and that the number of factors ranges between $1\leqslant q \leqslant q_{\mbox{true}} + 2$. These bounds are selected in order to speed up computation time without introducing any bias in the resulting inference (as confirmed by a smaller pilot study). For each scenario 500 datasets were simulated.

The main findings of the simulation study are illustrated in Figure \ref{fig:many}. Note that in Scenario 1 {\tt fabMix} almost always finds the correct clustering structure: the boxplots of the adjusted Rand Index are centered at 1 and, on the second row, the boxplots of the estimated number of clusters are centered at the corresponding true value. On the other hand, observe that for $K\geqslant 6$ {\tt pgmm} has the tendency to overestimate the number of clusters. In the more challenging Scenario 2 the estimates of the number of cluster exhibit larger variability. However note that for $K=8,9,10$ the number of clusters selected by {\tt fabMix} is closer to the true value than {\tt pgmm}, a fact which is also reflected in the ARI where {\tt fabMix} tends to have larger values than {\tt pgmm}. For both scenarios, the estimation of the number of factors is in strong agreement between the two methods, as shown in the third row of Figure \ref{fig:many}. In the last row, the selected parameterization is shown. Observe that the results are fairly consistent between the two methods. 

Finally, we note that in the presented simulation study, the generated clusters have equal sizes (on average). The reader is referred to Appendix C for exploring the performance of the compared methods in the presence of small and large clusters with respect to the size of the available data ($n$).

\subsection{Publicly available datasets}\label{sec:yeast}

\begin{figure*}[t]
\centering
\begin{tabular}{c}
\includegraphics[scale=0.5]{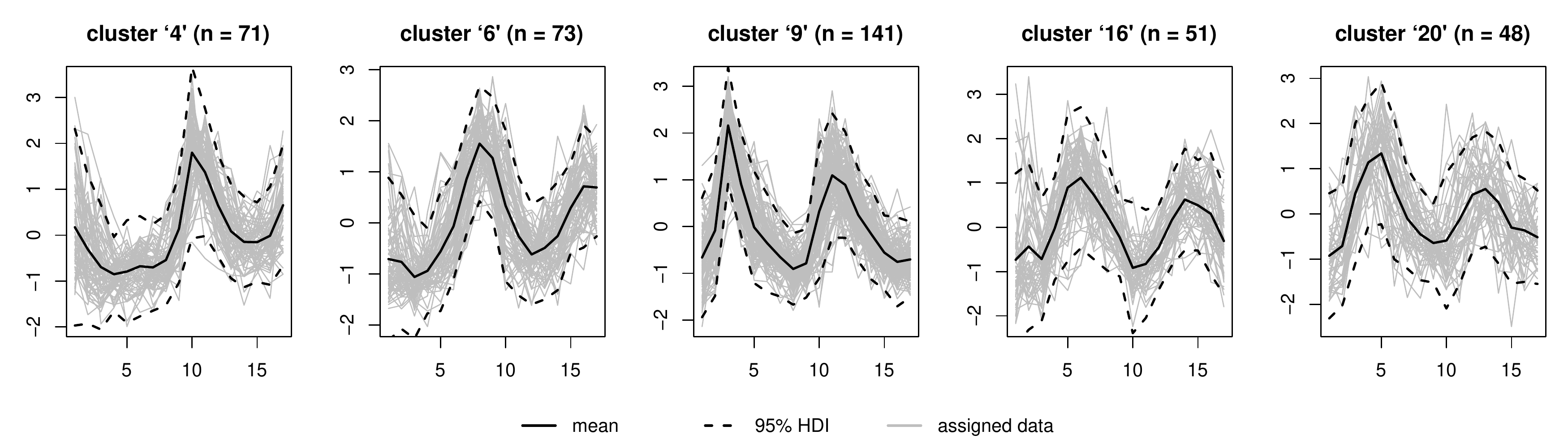}
\end{tabular}
\caption{Marginal mean with $95\%$ Highest Density Interval and the corresponding assigned data per alive cluster for the yeast cell cycle data.}
\label{fig:yeast}
\end{figure*}

\begin{table}[ht]
\centering
\caption{Selected number of clusters, factors, parameterization and adjusted rand index for the publicly available data.}
\begin{tabular}{p{14mm}p{2mm}p{2mm}p{6mm}p{5mm}p{2mm}p{2mm}p{6mm}p{5mm}}
\toprule
\multirow{2}{*}{Data (K)} &
\multicolumn{4}{c}{{\tt fabMix}} &
\multicolumn{4}{c}{{\tt pgmm}}\\
 \cline{2-9}\\[-0.8em]
 &  $\widehat{K}$ & $\widehat{q}$ & model & ARI &  $\widehat{K}$ & $\widehat{q}$ & model & ARI \\
\midrule 
Coffee (2) &  2 & 1 & CUU & 1 & 4 & 4 & CUU & .29\\
Wave (3) &   3&1 &UCU & .61 & 3 & 1 & UCU & .61\\
Wine (3) &  5 & 4 & CUU & .83 & 3 & 4 & CUU &.97\\
Yeast (5)&  5 & 6 & CUU & .50 & 20 & 10 &CUC & .20\\
\bottomrule
\end{tabular}
\label{tab:tab2}
\end{table}

In this section we analyze 4 publicly available datasets: a subset of the wave dataset \citep{wavedata, uci} available at the {\tt fabMix} package, the wine dataset \citep{forina1986multivariate} available at the {\tt pgmm} package, the coffee dataset \citep{Streuli} available at the {\tt pgmm} package, and the standardized yeast cell cycle data \citep{CHO199865} available at \url{http://faculty.washington.edu/kayee/model/}. Note that \cite{PAPASTAMOULIS2018220} analyzed the first three datasets but only considering the UUU and UCU parameterizations for {\tt fabMix}. 

The coffee dataset consists of $n=43$ coffee samples of $p = 12$ variables, collected from beans corresponding to the Arabica and Robusta species (thus, $K = 2$). The wave dataset consists of a randomly sampled subset of 1500 observations from the wave dataset \citep{wavedata}, available from the UCI machine learning repository \citep{uci}. According to the available ground-truth classification of the dataset, there are 3 equally weighted underlying classes of $21$-dimensional continuous data. The wine dataset \citep{forina1986multivariate}, available at the {\tt pgmm} package \citep{pgmm}, contains $p=27$ variables measuring chemical and physical properties of $n = 178$ wines, grouped in three types  (thus, $K = 3$). The reader is referred to  \cite{McNicholas2008, PAPASTAMOULIS2018220} for more detailed descriptions of the the data.

The yeast cell cycle data \citep{CHO199865} quantifies gene expression levels over two cell cycles (17 time points). The dataset has previously been used for evaluating the effectiveness of model-based clustering techniques \citep{doi:10.1093/bioinformatics/17.10.977}. We used the standardized subset of the 5-phase criterion, containing $n = 384$ genes measured at $p = 17$ time points. The expression levels of the $n=384$ genes  peak at different time points corresponding to the five phases of cell cycle, so this five class partition of the data is used as the ground-truth classification.

We applied our method using the 8 parameterizations of overfitting mixtures with {\tt Kmax = 20} components for $1\leqslant q\leqslant q_{\mbox{max}}$ factors using {\tt nChains = 4} heated chains. We set $q_{\mbox{max}} = 5$ for the coffee, wave and wine datasets, while $q_{\mbox{max}} = 10$ for the yeast cell cycle dataset.The number of MCMC cycles was set to {\tt mCycles = 1100}, while the first {\tt burnCycles = 100} were discarded as burn-in. The 8 parameterizations are processed in parallel on {\tt parallelModels = 4} cores, while each heated chain of a given parameterization is also running in parallel. All other prior parameters were fixed at their default values. 

We have also applied {\tt pgmm} considering the same range of clusters and factors per dataset. For each combination of number of factors, components and parameterization, the EM algorithm was initialized using 5 random starting values as well as the K-means clustering algorithm, that is, 6 different starts in total. For the coffee dataset a larger number of different starts is required as discussed in \cite{PAPASTAMOULIS2018220}.


Table \ref{tab:tab2} summarizes the results for each of the publicly available data. We conclude that {\tt fabMix} performs better than {\tt pgmm} at the coffee and yeast datasets. In the wine dataset, on the other hand, {\tt pgmm} performs better than {\tt fabMix}, but we underline the improved performance of our method compared to the one reported by \cite{PAPASTAMOULIS2018220} where only the UUU and UCU parameterizations were fitted. The two methods are in agreement on the wave dataset. The {\tt plot} command of the {\tt fabMix} package displays the estimated clusters according to the CUU model with 6 factors for the yeast dataset, as shown in Figure \ref{fig:yeast}.

\begin{figure*}[t]
\centering
\begin{tabular}{c}
\includegraphics[scale=0.55]{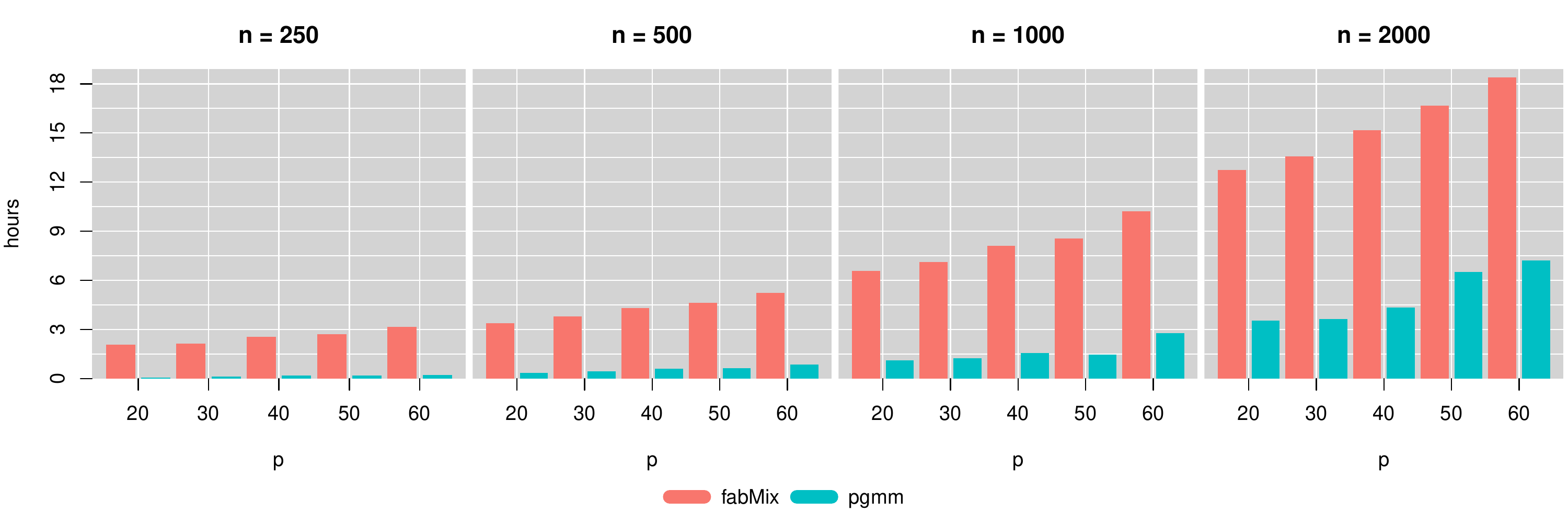}
\end{tabular}
\caption{Total time needed for fitting the 8 parameterizations considering $q =1,\ldots,5$ (40 models in total) for various levels of sample size ($n$) and number of variables ($p$). We considered  $K_{\mbox{max}} = 20$ components in {\tt fabMix} and $1\leqslant K\leqslant 20$ in {\tt pgmm}. Each parameterization is fitted in parallel using 8 threads. No multiple runs ({\tt pgmm}) or parallel chains ({\tt fabMix}) are considered. The MCMC algorithm in {\tt fabMix} ran for $12000$ iterations. The bars display averaged wallclock run times across 5 replicates.}
\label{fig:time}
\end{figure*}

\section{Discussion and further remarks}\label{sec:summary}

This study offered an efficient Bayesian methodology for  model-based clustering of multivariate data using mixtures of factor analyzers. The proposed model extended the ideas of \cite{PAPASTAMOULIS2018220} building upon the previously introduced set of parsimonious Gaussian mixture models \citep{McNicholas2008, mcnicholas2010serial}. The additional parameterizations improved the performance of the proposed method compared to \cite{PAPASTAMOULIS2018220} where only two out of eight parameterizations were available. Furthermore, our contributed {\tt R} package makes the proposed method available to a wider audience of researchers.

The computational cost of our MCMC method is larger than the EM algorithm, as shown in Figure \ref{fig:time}. But of course, when a point estimate is required, the EM algorithm is the quickest solution. When a point estimate is not sufficient, our method offers an attractive Bayesian treatment of the problem. Clearly, the Bayesian approach does show further advantages (as in the simulated datasets according to Scenario 1, as well as in the coffee and yeast datasets), where the multimodality of the likelihood potentially causes the EM to converge to local maxima. 

A direction for future research is to generalize the method in order to automatically detect the number of factors in a fully Bayesian manner. This is possible by e.g.~treating the number of factors as a random variable and implementing a reversible jump mechanism in order to update it inside the MCMC sampler. Another possibility would be to incorporate strategies for searching the space of sparse factor loading matrices allowing posterior inference for factor selection \citep{bhattacharya2011sparse, doi:10.1111/bmsp.12019, conti2014bayesian}.  Recent advances on infinite mixtures  of infinite factor models \citep{murphy2017infinite} also allow for direct inference of the number of clusters and factors and could boost the flexibility of our modelling approach.

\begin{acknowledgements}
The author would like to acknowledge the assistance given by IT services and use of the Computational Shared Facility of the University of Manchester. The suggestions of two anonymous reviewers helped to improve the findings of this study.
\end{acknowledgements}

\appendix
\renewcommand{\theequation}{\thesection.\arabic{equation}}
\renewcommand\thefigure{\thesection.\arabic{figure}}
\renewcommand\thetable{\thesection.\arabic{table}}

\section{Overfitted mixture model}\label{sec:overfitting}

Assume that the observed data has been generated from a mixture model with $K_0$ components $$f_{K_0}(\bs x) = \sum_{k=1}^{K_0}w_kf_k(\bs x|\bs\theta_k),$$ where $f_k\in\mathcal F_{\Theta}=\{f(\cdot|\bs\theta):\bs\theta\in\Theta\}$; $k = 1,\ldots,K_0$ denotes a member of a parametric family of distributions. Consider that an overfitted mixture model $f_{K}(\bs x)$ with $K > K_0$ components is fitted to the data. \cite{rousseau2011asymptotic} showed  that the asymptotic behaviour of the posterior distribution of the $K-K_0$ redundant components depends on the prior distribution of mixing proportions $(\bs w)$. Let $d$ denote the dimension of free parameters of the  distribution $f_k$. For the case of a Dirichlet prior distribution, 
\begin{equation}\label{eq:dirichlet_prior}
\bs w\sim\mathcal D\left(\gamma_1,\ldots,\gamma_K\right)
\end{equation}
if $$\max\{\gamma_k; k=1,\ldots,K\} < d/2$$ then the posterior weight of the extra components converges to zero (Theorem 1 of \cite{rousseau2011asymptotic}). 

Let $f_K(\bs\theta,\bs z|\bs x)$ denote the joint posterior distribution of model parameters and latent allocation variables for a model with $K$ components. When using an overfitted mixture model, the inference on the number of clusters reduces to (a): choosing a sufficiently large value of mixture components ($K$), (b): running a typical MCMC sampler for drawing samples from the posterior distribution $f_K(\bs\theta, \bs z|\bs x)$ and (c) inferring the number of ``alive'' mixture components.  Note that at MCMC iteration $t = 1,2,\ldots$ (c) reduces to keeping track of the number of elements in the set $\boldsymbol{K_0}^{(t)}=\{k=1,\ldots,K: \sum_{i=1}^{n}I(z_i^{(t)}=k)>0\}$, where $z_i^{(t)}$ denotes the simulated allocation of observation $i$ at iteration $t$.

In our case the dimension of free parameters in the $k$-th mixture component is equal to $d = 2p + pq-\frac{q(q-1)}{2}$. Following \cite{PAPASTAMOULIS2018220}, we set $\gamma_1=\ldots=\gamma_K = \frac{\gamma}{K}$, thus the distribution of mixing proportions in Equation \eqref{eq:dirichlet_prior} becomes
\begin{equation}\label{eq:dirichlet_prior_same}
 \bs w\sim\mathcal D\left(\frac{\gamma}{K},\ldots,\frac{\gamma}{K}\right)
\end{equation}
where $0 < \gamma < d/2$ denotes a pre-specified positive number. Such a value is chosen for two reasons. At first, it is smaller than $d/2$ so the asymptotic results of \cite{rousseau2011asymptotic} ensure that extra components will be emptied as $n\rightarrow\infty$. Second, this choice can be related to standard practice when using Bayesian non-parametric clustering methods where the parameters of a mixture are drawn from a Dirichlet process \citep{ferguson}, that is, a Dirichlet process mixture model \citep{neal2000markov}.

\section{Details of the MCMC sampler}\label{sec:mcmc_details}

\textbf{Data normalization and prior parameters} Before running the sampler, the raw data is standardized by applying the $z$-transformation
\begin{equation*}
\frac{x_{ir} - \bar x_{r}}{\sqrt{s^2_r}}, \quad i =1,\ldots,n;r = 1,\ldots,p
\end{equation*}
where $\bar x_{r} = \frac{\sum_{i=1}^{n}x_{ir}}{n}$ and $s^{2}_{r}= \frac{1}{n-1}\sum_{i=1}^{n}\left(x_{ir}-\bar x_{r}\right)^2$. The main reason for using standardized data is that the sampler mixes better. Furthermore, it is easier to choose prior parameters that are not depending on the observed data, that is, using the data twice. In any other case, one could use empirical prior distributions as reported in \cite{Fokoue2003}, see also \cite{dellaportas2006multivariate}. For the case of standardized data, the prior parameters are specified in Table \ref{tab:prior}. Standardized data is also used as input to {\tt pgmm}. 

\begin{table}[!ht]
\centering
\caption{Prior parameter specification for the case of standardized data.}

\begin{tabular}{llllllll}
\toprule
 & $\alpha$& $\beta$& $\gamma$& $g$& $h$& $\bs\xi = (\xi_1,\ldots,\xi_p)^T$& $\bs\Psi$\\
  \midrule
value & 0.5  & 0.5 & 1 &  0.5 & 0.5 & $(0,\ldots,0)^T$ &  $\bs{\mathrm{I}}_p$ \\ \bottomrule
\end{tabular}
\label{tab:prior}
\end{table}

\textbf{Prior parallel tempering} It is well known that the posterior surface of mixture models can exhibit many local modes \citep{celeux2000computational, Marin:05}. In such cases simple MCMC algorithms may become trapped in minor modes
 and demand a very large number of iterations to sufficiently explore the posterior distribution. In order to produce a well-mixing MCMC sample and improve the convergence of our algorithm we utilize ideas from parallel tempering schemes \cite{geyer1991, geyer1995, Altekar12022004}, where different chains are running in parallel and they are allowed to switch states. Each chain corresponds to a different  posterior distribution, and usually each one represents a ``heated'' version of the target posterior distribution. This is achieved by raising the original target to a power $T$ with $0\leqslant T \leqslant 1$, which flattens the posterior surface, thus, easier to explore when using an MCMC sampler. 

In the context of overfitting mixture models, \cite{overfitting}  introduced a prior parallel tempering scheme, which  is also applied by  \cite{PAPASTAMOULIS2018220}. Under this approach, each heated chain corresponds to a model with identical likelihood as the original, but with a different prior distribution. Although the prior tempering can be imposed on any subset of parameters, it is only applied to the Dirichlet prior distribution of mixing proportions \citep{overfitting}. 
Let us denote by $f_i(\bs\varphi|\bs x)$ and $f_i(\bs\varphi)$; $i=1,\ldots,J$, the posterior and prior distribution of the $i$-th chain, respectively. Obviously, $f_i(\bs\varphi|\bs x) \propto f(\bs x|\bs\varphi)f_i(\bs\varphi)$. Let $\bs\varphi^{(t)}_i$ denote the state of chain $i$ at iteration $t$ and assume that a swap between chains $i$ and $j$ is proposed. The proposed move is accepted with probability $\min\{1,A\}$ where
\begin{equation}\label{eq:mh_ar}A = \frac{f_i(\bs\varphi_j^{(t)}|\bs x)f_j(\bs\varphi_i^{(t)}|\bs x)}{f_i(\bs\varphi_i^{(t)}|\bs x)f_j(\bs\varphi_j^{(t)}|\bs x)}=
\frac{f_i(\bs\varphi_j^{(t)})f_j(\bs\varphi_i^{(t)})}{f_i(\bs\varphi_i^{(t)})f_j(\bs\varphi_j^{(t)})}=
\frac{\widetilde{f}_i( w_j^{(t)})\widetilde{f}_j( w_i^{(t)})}{\widetilde{f}_i( w_i^{(t)})\widetilde{f}_j( w_j^{(t)})},\end{equation}
and $\widetilde{f}_i(\cdot)$ corresponds to the probability density function of the Dirichlet prior distribution related to chain $i = 1,\ldots,J$. According to Equation \eqref{eq:dirichlet_prior_same}, this is
\begin{equation}\label{eq:dir_prior_tempering}
 \bs w\sim\mathcal D\left(\frac{\gamma_{(j)}}{K},\ldots,\frac{\gamma_{(j)}}{K}\right),
\end{equation} 
for a pre-specified set of parameters $\gamma_{(j)}>0$ for $j = 1,\ldots,J$.

In our examples we used a total of $J = 4$ parallel chains where the prior distribution of mixing proportions for chain $j$ in Equation \eqref{eq:dir_prior_tempering} is selected as
\begin{equation*}
\gamma_{(j)} = \gamma + \delta(j-1), \quad j = 1,\ldots,J,
\end{equation*}
where $\delta > 0$. For example, when the overfitting mixture model uses $K=20$ components and  $\gamma = 1$ (the default value shown in Table \ref{tab:prior}), it follows from Equation \eqref{eq:dirichlet_prior_same} that the parameter vector of the Dirichlet prior of mixture weights which corresponds to the target posterior distribution ($j = 1$) is equal to $(0.05,\ldots,0.05)$. Also in our examples we have used $\delta =1$, but in general we strongly suggest to tune this parameter until a reasonable acceptance rate is achieved. Each chain runs in parallel and every 10 iterations we randomly select two adjacent chains $(j,j+1)$, $j\in\{1,\ldots,J-1\}$ and propose to swap their current states. A proposed swap is accepted with probability $A$ in Equation \eqref{eq:mh_ar}.

\textbf{``Overfitting initialization'' strategy} We briefly describe the ``overfitting initialization'' procedure introduced by \cite{PAPASTAMOULIS2018220}.
 We used an initial period of 500 MCMC iterations where each chain is initialized from totally random starting values, but under a Dirichlet prior distribution  with large prior parameter values. These values were chosen in a way that the asymptotic results of \cite{rousseau2011asymptotic} guarantee that the redundant mixture components will have non-negligible posterior weights. More specifically for chain $j$ we assume $ w\sim\mathcal D(\gamma'_{j},\ldots,\gamma'_{j})$ with $\gamma'_{(j)} = \frac{d}{2} + (j-1)\frac{d}{2(J-1)}$, for $j =1,\ldots,J$.  Then, we initialize the actual model by this state. According to  \cite{PAPASTAMOULIS2018220},  this specific scheme was found to outperform other initialization procedures.


\begin{table*}[t]
\centering
\caption{Setup for  our simulation scenarios with unequal cluster sizes. In all cases the dimensionality of the multivariate data is equal to $p=30$.}

\begin{tabular}{lll}
\toprule
& Scenario 1 & Scenario 2\\
\midrule
True number of clusters $(K)$ & $5$& $2$\\
True number of factors $(q)$& $2$& $3$\\
True mixing proportions $(\bs w)$& $\left(\frac{1}{15},\frac{2}{15},\frac{3}{15},\frac{4}{15},\frac{5}{15}\right)$& $ \left(\frac{1}{20},\frac{19}{20}\right)$\\
Sample size ($n$) & $50, 100, 200, 300, 400, 500$ & $50, 100, 200,300, \ldots,1500$\\
\bottomrule
\end{tabular}
\label{tab:small}
\end{table*}

\begin{figure*}[ht]
\centering
\includegraphics[scale=0.7]{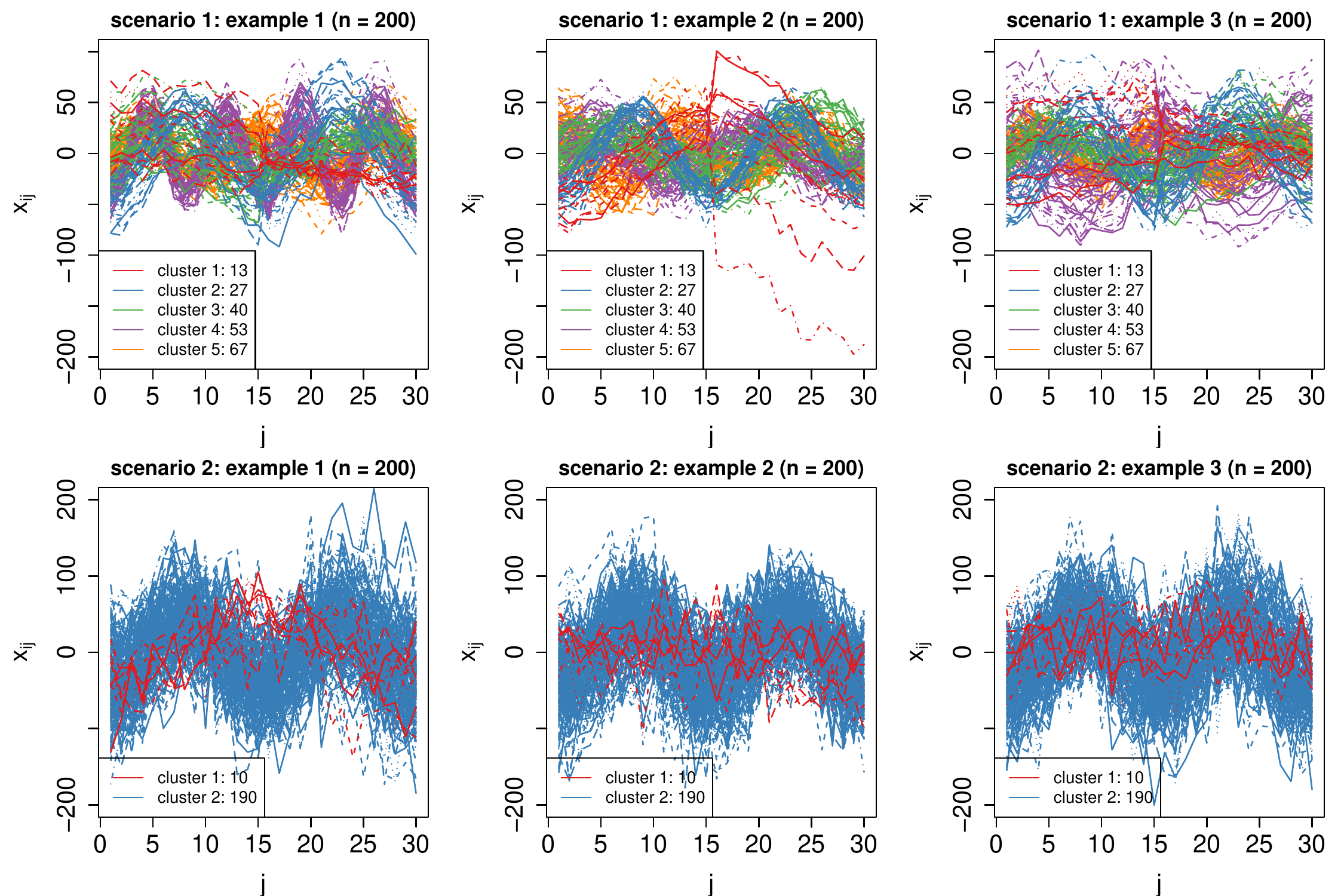}
\caption{Examples of simulated datasets with unequal cluster sizes according to scenarios 1 and 2 and $n = 200$. The legend  shows the ``true'' cluster sizes.}
\label{fig:small}
\end{figure*}

\section{Additional simulations}\label{sec:more_sim}

In the simulation section of the manuscript,  the weights of the simulated datasets have been randomly generated from a Dirichlet distribution with mean equal to $1/K$, conditional on the number of clusters ($K$). Thus, on average, the true cluster sizes are equal. In this section we examine the performance of the proposed method in the presence of unequal cluster sizes with respect to the size ($n$) of the observed data.


 We replicate the simulation mechanism for scenarios 1 and 2 presented in the main text, but now we consider unequal (true) cluster sizes, as detailed in Table \ref{tab:small}. For each case, the sample size is increasing (as shown in the last column of Table \ref{tab:small}) while keeping all others parameters (that is, the true values of marginal means and factor loadings) constant. As shown in Table \ref{tab:small}, in scenario 1 there are $5$ clusters and $2$ factors, whereas in scenario 2 there are $2$ clusters and $3$ factors. In total 3 different examples per scenario are considered: for a given scenario, the component-specific parameters are different in each example but the weights are the same. An instance of our three examples (per scenario) using $n = 200$ simulated observations is shown at Figure \ref{fig:small}. Observe that in all cases the ``true clusters'' are not easily distinguishable, especially in scenario 2 where there is a high degree of cluster overlapping.

\begin{figure*}[ht]
\centering
\includegraphics[scale=0.7]{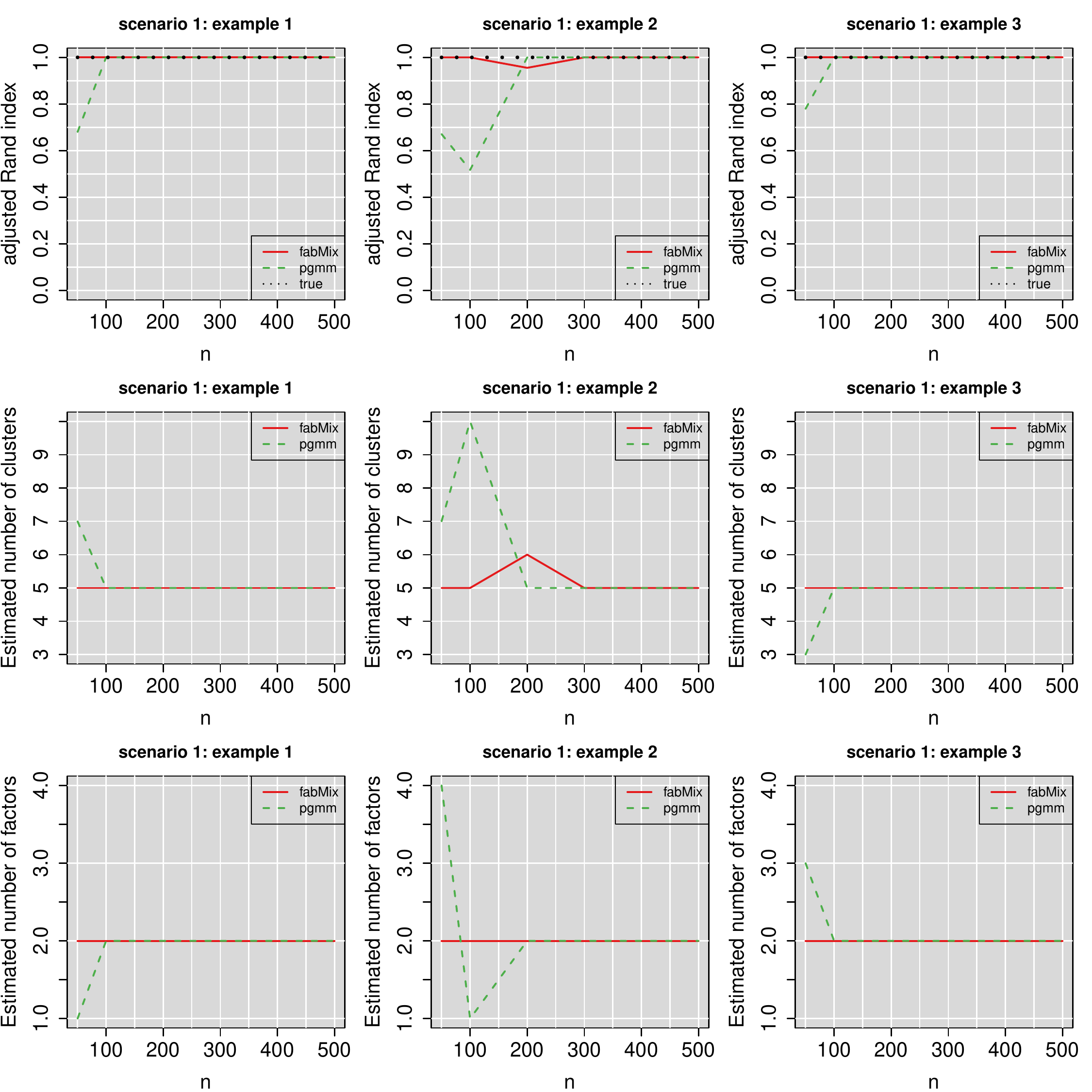}
\caption{Adjusted Rand index (1st row), estimated number of clusters (2nd row) and estimated number of factors (3rd row) for simulated data according to scenario 1 with unequal cluster sizes and increasing sample size. The dotted line in the first row corresponds to the adjusted Rand index between the ground-truth and the clustering of the data when applying the Maximum A Posteriori rule using the parameter values that generated the data \eqref{eq:map}. For all examples, the true number of clusters and factors is equal to $5$ and $2$, respectively. }
\label{fig:many1}
\end{figure*}

\begin{figure*}[ht]
\centering
\includegraphics[scale=0.7]{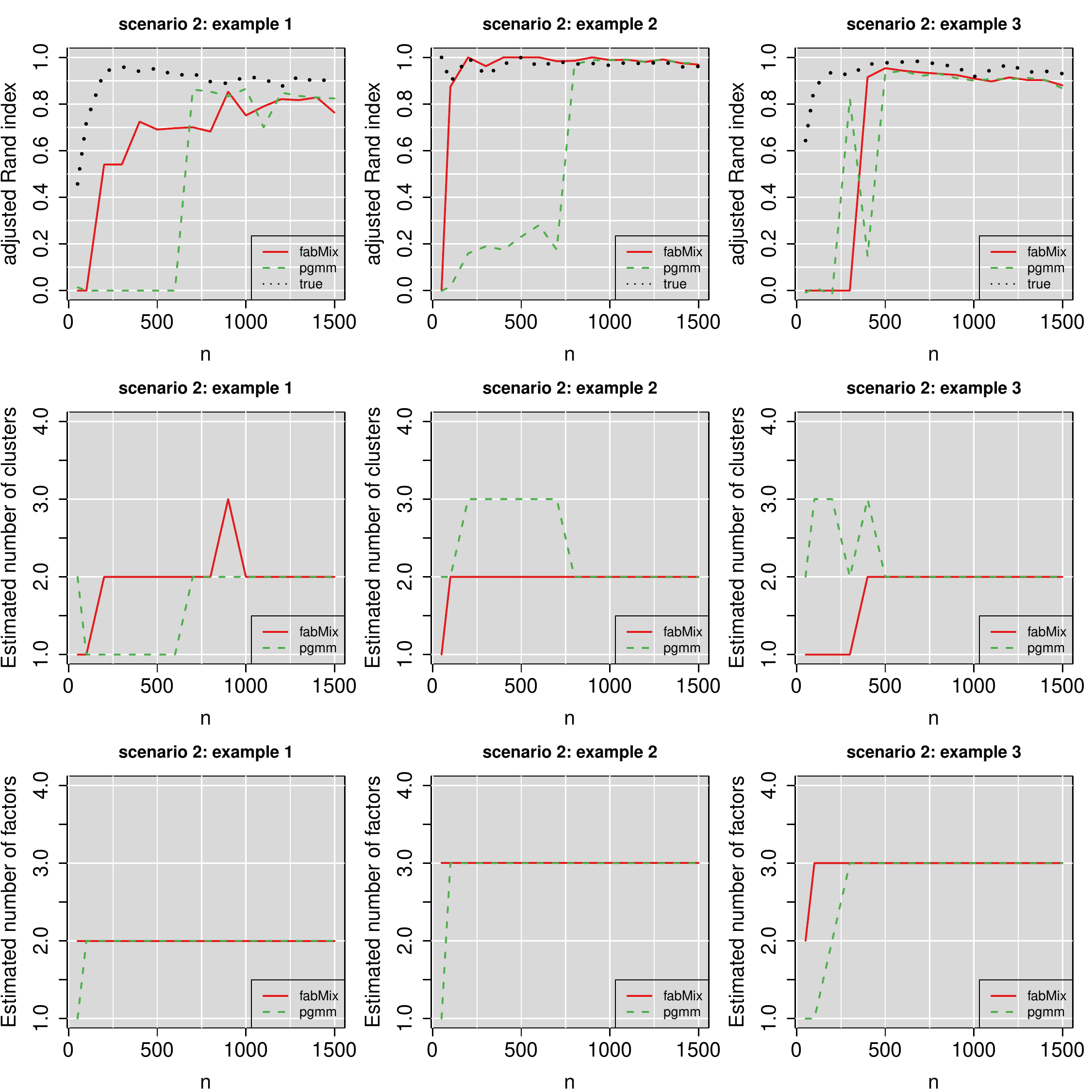}
\caption{Adjusted Rand index (1st row), estimated number of clusters (2nd row) and estimated number of factors (3rd row) for simulated data according to scenario 2 with unequal cluster sizes and increasing sample size. The dotted line in the first row corresponds to the adjusted Rand index between the ground-truth and the clustering of the data when applying the Maximum A Posteriori rule using the parameter values that generated the data \eqref{eq:map}. For all examples, the true number of clusters and factors is equal to $2$ and $3$, respectively. }
\label{fig:many2}
\end{figure*}

We applied {\tt fabMix} and {\tt pgmm} using the same number (4) of parallel chains (for {\tt fabMix}) and different starts (for {\tt pgmm}) as in the simulations presented in the main paper. The results are summarized in Figures \ref{fig:many1} and \ref{fig:many2} for scenarios 1 and 2, respectively. The adjusted Rand  Index is displayed in the first line of each Figure, where the horizontal axis denotes the sample size ($n$) of the synthetic data. The dotted black line corresponds to the adjusted Rand index between the ground-truth and the cluster assignments arising when applying the Maximum  A Posteriori rule using the true parameter values that generated the data, that is,
\begin{equation}\label{eq:map}
{z}_i=\max_{k\in\{1,\ldots,K^*\}}\left\{\frac{w_k^*f_k(\bs x_i|\theta_k^*)}{\sum_{j=1}^{K^*}w_j^*f_j(\bs x_i|\theta_j^*)}\right\},\quad i=1,\ldots,n
\end{equation}
where $K^*$, $(w_1^*,\ldots,w^*_{K^*})$ and $(\theta_1^*,\ldots,\theta^*_{K^*})$ denote the values of number of components, mixing proportions and parameters of the multivariate normal densities of the mixture model used to generate the data. Observe that in all three examples of   Scenario 1 the dotted black line  is always equal to 1, but this is not the case in the more challenging Scenario  2 due to enhanced levels of cluster overlapping.

The adjusted Rand index between the ground truth clustering and the estimated cluster assignments arising from {\tt fabMix} and {\tt pgmm} are shown in the the first row of Figures  \ref{fig:many1} and \ref{fig:many2}. Clearly, the compared methods have similar performance as the sample  size increases, but for smaller values of $n$ the proposed method outperforms the {\tt pgmm} package. 

The estimated number of clusters, shown at the second row of Figures  \ref{fig:many1} and \ref{fig:many2} agree (in most cases) with the true number of clusters, but note that our method is capable of detecting the right value earlier than {\tt pgmm}. Two exceptions occur at $n=200$ for example 2 of scenario 1 where {\tt fabMix} (red line at 2nd row of Figure \ref{fig:many1}) inferred 6 alive clusters instead of 5, as well as at $n=800$ for example 1 of scenario 2 where {\tt fabMix} (red line at 2nd row of Figure \ref{fig:many2}) inferred 3 alive clusters instead of 2. 

Finally, the last row in Figures \ref{fig:many1} and \ref{fig:many2} displays the inferred number of factors for Scenarios 1 and 2, respectively. In every single case, the estimate arising from {\tt fabMix} is at least as close as the estimate arising from {\tt pgmm} to the corresponding true value. Note however that in example 1 of scenario 2 both methods detect a smaller number of factors (2 instead of 3 factors). In all other cases we observe that as the sample size increases both methods infer the ``true'' number of factors.

\bibliographystyle{spbasic}      
\bibliography{papastamoulis}   

\begin{thebibliography}{65}
\providecommand{\natexlab}[1]{#1}
\providecommand{\url}[1]{{#1}}
\providecommand{\urlprefix}{URL }
\expandafter\ifx\csname urlstyle\endcsname\relax
  \providecommand{\doi}[1]{DOI~\discretionary{}{}{}#1}\else
  \providecommand{\doi}{DOI~\discretionary{}{}{}\begingroup
  \urlstyle{rm}\Url}\fi
\providecommand{\eprint}[2][]{\url{#2}}

\bibitem[{Altekar et~al.(2004)Altekar, Dwarkadas, Huelsenbeck, and
  Ronquist}]{Altekar12022004}
Altekar G, Dwarkadas S, Huelsenbeck JP, Ronquist F (2004) Parallel {M}etropolis
  coupled {M}arkov chain {M}onte {C}arlo for {B}ayesian phylogenetic inference.
  Bioinformatics 20(3):407--415, \doi{10.1093/bioinformatics/btg427},
  \urlprefix\url{http://bioinformatics.oxfordjournals.org/content/20/3/407.abstract},
  \eprint{http://bioinformatics.oxfordjournals.org/content/20/3/407.full.pdf+html}

\bibitem[{Bartholomew et~al.(2011)Bartholomew, Knott, and
  Moustaki}]{bartholomew2011latent}
Bartholomew DJ, Knott M, Moustaki I (2011) Latent variable models and factor
  analysis: A unified approach, vol 904. John Wiley \& Sons

\bibitem[{Bhattacharya and Dunson(2011)}]{bhattacharya2011sparse}
Bhattacharya A, Dunson DB (2011) Sparse {B}ayesian infinite factor models.
  Biometrika pp 291--306

\bibitem[{Breiman et~al.(1984)Breiman, Friedman, Olshen, and Stone}]{wavedata}
Breiman L, Friedman J, Olshen R, Stone C (1984) Classification and Regression
  Trees. Wadsworth International Group: Belmont, California

\bibitem[{Celeux et~al.(2000{\natexlab{a}})Celeux, Hurn, and
  Robert}]{doi:10.1080/01621459.2000.10474285}
Celeux G, Hurn M, Robert CP (2000{\natexlab{a}}) Computational and inferential
  difficulties with mixture posterior distributions. Journal of the American
  Statistical Association 95(451):957--970,
  \doi{10.1080/01621459.2000.10474285},
  \urlprefix\url{https://amstat.tandfonline.com/doi/abs/10.1080/01621459.2000.10474285},
  \eprint{https://amstat.tandfonline.com/doi/pdf/10.1080/01621459.2000.10474285}

\bibitem[{Celeux et~al.(2000{\natexlab{b}})Celeux, Hurn, and
  Robert}]{celeux2000computational}
Celeux G, Hurn M, Robert CP (2000{\natexlab{b}}) Computational and inferential
  difficulties with mixture posterior distributions. Journal of the American
  Statistical Association 95(451):957--970

\bibitem[{Cho et~al.(1998)Cho, Campbell, Winzeler, Steinmetz, Conway, Wodicka,
  Wolfsberg, Gabrielian, Landsman, Lockhart, and Davis}]{CHO199865}
Cho RJ, Campbell MJ, Winzeler EA, Steinmetz L, Conway A, Wodicka L, Wolfsberg
  TG, Gabrielian AE, Landsman D, Lockhart DJ, Davis RW (1998) A genome-wide
  transcriptional analysis of the mitotic cell cycle. Molecular Cell 2(1):65 --
  73, \doi{https://doi.org/10.1016/S1097-2765(00)80114-8},
  \urlprefix\url{http://www.sciencedirect.com/science/article/pii/S1097276500801148}

\bibitem[{Conti et~al.(2014)Conti, Fr{\"u}hwirth-Schnatter, Heckman, and
  Piatek}]{conti2014bayesian}
Conti G, Fr{\"u}hwirth-Schnatter S, Heckman JJ, Piatek R (2014) Bayesian
  exploratory factor analysis. Journal of econometrics 183(1):31--57

\bibitem[{Dellaportas and Papageorgiou(2006)}]{dellaportas2006multivariate}
Dellaportas P, Papageorgiou I (2006) Multivariate mixtures of normals with
  unknown number of components. Statistics and Computing 16(1):57--68

\bibitem[{Dempster et~al.(1977)Dempster, Laird, and Rubin}]{Dempster:77}
Dempster AP, Laird NM, Rubin D (1977) Maximum {L}ikelihood from {I}ncomplete
  {D}ata via the {EM} {A}lgorithm (with discussion). Journal of the Royal
  Statistical Society B 39:1--38

\bibitem[{Eddelbuettel and Fran\c{c}ois(2011)}]{rcpp}
Eddelbuettel D, Fran\c{c}ois R (2011) {Rcpp}: Seamless {R} and {C++}
  integration. Journal of Statistical Software 40(8):1--18,
  \doi{10.18637/jss.v040.i08},
  \urlprefix\url{http://www.jstatsoft.org/v40/i08/}

\bibitem[{Eddelbuettel and Sanderson(2014)}]{rcpparmadillo}
Eddelbuettel D, Sanderson C (2014) Rcpparmadillo: Accelerating {R} with
  high-performance {C++} linear algebra. Computational Statistics and Data
  Analysis 71:1054--1063,
  \urlprefix\url{http://dx.doi.org/10.1016/j.csda.2013.02.005}

\bibitem[{Ferguson(1973)}]{ferguson}
Ferguson TS (1973) A {B}ayesian analysis of some nonparametric problems. The
  Annals of Statistics 1(2):209--230

\bibitem[{Fokou{\'e} and Titterington(2003)}]{Fokoue2003}
Fokou{\'e} E, Titterington D (2003) Mixtures of factor analysers. {B}ayesian
  estimation and inference by stochastic simulation. Machine Learning
  50(1):73--94

\bibitem[{Forina et~al.(1986)Forina, Armanino, Castino, and
  Ubigli}]{forina1986multivariate}
Forina M, Armanino C, Castino M, Ubigli M (1986) Multivariate data analysis as
  a discriminating method of the origin of wines. Vitis 25(3):189--201

\bibitem[{Fraley and Raftery(2002)}]{mclust1}
Fraley C, Raftery AE (2002) Model-based clustering, discriminant analysis and
  density estimation. Journal of the American Statistical Association
  97:611--631

\bibitem[{Fraley et~al.(2012)Fraley, Raftery, Murphy, and Scrucca}]{mclust2}
Fraley C, Raftery AE, Murphy TB, Scrucca L (2012) mclust Version 4 for R:
  Normal Mixture Modeling for Model-Based Clustering, Classification, and
  Density Estimation

\bibitem[{Gaujoux(2018)}]{dorng}
Gaujoux R (2018) doRNG: Generic Reproducible Parallel Backend for 'foreach'
  Loops. \urlprefix\url{https://CRAN.R-project.org/package=doRNG}, r package
  version 1.7.1

\bibitem[{Gelfand and Smith(1990)}]{gelfand}
Gelfand A, Smith A (1990) Sampling-based approaches to calculating marginal
  densities. Journal of American Statistical Association 85:398--409

\bibitem[{Geman and Geman(1984)}]{geman}
Geman S, Geman D (1984) Stochastic relaxation, gibbs distributions, and the
  {B}ayesian restoration of images. IEEE Transactions on Pattern Analysis and
  Machine Intelligence PAMI-6(6):721--741, \doi{10.1109/TPAMI.1984.4767596}

\bibitem[{Geweke and Zhou(1996)}]{doi:10.1093/rfs/9.2.557}
Geweke J, Zhou G (1996) Measuring the pricing error of the arbitrage pricing
  theory. The Review of Financial Studies 9(2):557--587,
  \doi{10.1093/rfs/9.2.557},
  \urlprefix\url{http://dx.doi.org/10.1093/rfs/9.2.557},
  \eprint{/oup/backfile/content_public/journal/rfs/9/2/10.1093_rfs_9.2.557/1/090557.pdf}

\bibitem[{Geyer(1991)}]{geyer1991}
Geyer CJ (1991) Markov chain {M}onte {C}arlo maximum likelihood. In:
  Proceedings of the 23rd Symposium on the Interface, Interface Foundation,
  Fairfax Station, Va, pp 156--163

\bibitem[{Geyer and Thompson(1995)}]{geyer1995}
Geyer CJ, Thompson EA (1995) Annealing {M}arkov chain {M}onte {C}arlo with
  applications to ancestral inference. Journal of the American Statistical
  Association 90(431):909--920, \doi{10.1080/01621459.1995.10476590},
  \urlprefix\url{http://www.tandfonline.com/doi/abs/10.1080/01621459.1995.10476590},
  \eprint{http://www.tandfonline.com/doi/pdf/10.1080/01621459.1995.10476590}

\bibitem[{Ghahramani et~al.(1996)Ghahramani, Hinton
  et~al.}]{ghahramani1996algorithm}
Ghahramani Z, Hinton GE, et~al. (1996) The em algorithm for mixtures of factor
  analyzers. Tech. rep., Technical Report CRG-TR-96-1, University of Toronto

\bibitem[{Green(1995)}]{Green:95}
Green PJ (1995) Reversible jump {M}arkov chain {M}onte {C}arlo computation and
  {B}ayesian model determination. Biometrika 82(4):711--732

\bibitem[{Hager(1989)}]{10.2307/2030425}
Hager WW (1989) Updating the inverse of a matrix. SIAM Review 31(2):221--239

\bibitem[{van Havre et~al.(2015)van Havre, White, Rousseau, and
  Mengersen}]{overfitting}
van Havre Z, White N, Rousseau J, Mengersen K (2015) Overfitting {B}ayesian
  mixture models with an unknown number of components. PLOS ONE 10(7):1--27

\bibitem[{Ihaka and Gentleman(1996)}]{ihaka:1996}
Ihaka R, Gentleman R (1996) R: A language for data analysis and graphics.
  Journal of Computational and Graphical Statistics 5(3):299--314,
  \urlprefix\url{https://doi.org/10.1080/10618600.1996.10474713}

\bibitem[{Kim and Mueller(1978)}]{kim1978factor}
Kim JO, Mueller CW (1978) Factor analysis: Statistical methods and practical
  issues, vol~14. Sage

\bibitem[{Ledermann(1937)}]{ledermann1937rank}
Ledermann W (1937) On the rank of the reduced correlational matrix in
  multiple-factor analysis. Psychometrika 2(2):85--93

\bibitem[{Lichman(2013)}]{uci}
Lichman M (2013) {UCI} machine learning repository.
  \urlprefix\url{http://archive.ics.uci.edu/ml}

\bibitem[{Marin et~al.(2005)Marin, Mengersen, and Robert}]{Marin:05}
Marin J, Mengersen K, Robert C (2005) {B}ayesian modelling and inference on
  mixtures of distributions. Handbook of Statistics 25(1):577--590

\bibitem[{Mavridis and Ntzoufras(2014)}]{doi:10.1111/bmsp.12019}
Mavridis D, Ntzoufras I (2014) Stochastic search item selection for factor
  analytic models. British Journal of Mathematical and Statistical Psychology
  67(2):284--303, \doi{10.1111/bmsp.12019},
  \urlprefix\url{https://onlinelibrary.wiley.com/doi/abs/10.1111/bmsp.12019},
  \eprint{https://onlinelibrary.wiley.com/doi/pdf/10.1111/bmsp.12019}

\bibitem[{McLachlan and Peel(2000)}]{McLachlan:00}
McLachlan J, Peel D (2000) Finite {M}ixture {M}odels. Wiley, New York

\bibitem[{McNicholas(2016)}]{mcnicholas2016mixture}
McNicholas PD (2016) Mixture model-based classification. CRC Press

\bibitem[{McNicholas and Murphy(2008)}]{McNicholas2008}
McNicholas PD, Murphy TB (2008) Parsimonious {G}aussian mixture models.
  Statistics and Computing 18(3):285--296

\bibitem[{McNicholas and Murphy(2010)}]{doi:10.1093/bioinformatics/btq498}
McNicholas PD, Murphy TB (2010) Model-based clustering of microarray expression
  data via latent {G}aussian mixture models. Bioinformatics 26(21):2705

\bibitem[{McNicholas et~al.(2010)McNicholas, Murphy, McDaid, and
  Frost}]{mcnicholas2010serial}
McNicholas PD, Murphy TB, McDaid AF, Frost D (2010) Serial and parallel
  implementations of model-based clustering via parsimonious {G}aussian mixture
  models. Computational Statistics \& Data Analysis 54(3):711--723

\bibitem[{McNicholas et~al.(2015)McNicholas, ElSherbiny, Jampani, McDaid,
  Murphy, and Banks}]{pgmm}
McNicholas PD, ElSherbiny A, Jampani RK, McDaid AF, Murphy B, Banks L (2015)
  pgmm: Parsimonious {G}aussian Mixture Models.
  \urlprefix\url{http://CRAN.R-project.org/package=pgmm}, {R} package version
  1.2.3

\bibitem[{McParland et~al.(2017)McParland, Phillips, Brennan, Roche, and
  Gormley}]{mcparland2017clustering}
McParland D, Phillips CM, Brennan L, Roche HM, Gormley IC (2017) Clustering
  high-dimensional mixed data to uncover sub-phenotypes: joint analysis of
  phenotypic and genotypic data. Statistics in medicine 36(28):4548--4569

\bibitem[{Meng and Van~Dyk(1997)}]{RSSB:RSSB082}
Meng XL, Van~Dyk D (1997) The {E}{M} algorithm -- an old folk-song sung to a
  fast new tune. Journal of the Royal Statistical Society: Series B
  (Statistical Methodology) 59(3):511--567

\bibitem[{Murphy et~al.(2019)Murphy, Gormley, and Viroli}]{murphy2017infinite}
Murphy K, Gormley IC, Viroli C (2019) Infinite mixtures of infinite factor
  analysers. arXiv preprint arXiv:170107010

\bibitem[{Neal(2000)}]{neal2000markov}
Neal RM (2000) Markov chain sampling methods for {D}irichlet process mixture
  models. Journal of computational and graphical statistics 9(2):249--265

\bibitem[{Nobile and Fearnside(2007)}]{Nobile2007}
Nobile A, Fearnside AT (2007) {B}ayesian finite mixtures with an unknown number
  of components: The allocation sampler. Statistics and Computing
  17(2):147--162, \doi{10.1007/s11222-006-9014-7},
  \urlprefix\url{http://dx.doi.org/10.1007/s11222-006-9014-7}

\bibitem[{Papastamoulis(2014)}]{papastamoulis2014handling}
Papastamoulis P (2014) Handling the label switching problem in latent class
  models via the {E}{C}{R} algorithm. Communications in Statistics-Simulation
  and Computation 43(4):913--927

\bibitem[{Papastamoulis(2016)}]{papastamoulis2016label}
Papastamoulis P (2016) label.switching: An {R} package for dealing with the
  label switching problem in {M}{C}{M}{C} outputs. Journal of Statistical
  Software 69(1):1--24

\bibitem[{Papastamoulis(2018{\natexlab{a}})}]{fabMix}
Papastamoulis P (2018{\natexlab{a}}) fabMix: Overfitting Bayesian Mixtures of
  Factor Analyzers with Parsimonious Covariance and Unknown Number of
  Components. \urlprefix\url{http://CRAN.R-project.org/package=fabMix}, {R}
  package version 4.5

\bibitem[{Papastamoulis(2018{\natexlab{b}})}]{PAPASTAMOULIS2018220}
Papastamoulis P (2018{\natexlab{b}}) Overfitting {B}ayesian mixtures of factor
  analyzers with an unknown number of components. Computational Statistics \&
  Data Analysis 124:220 -- 234,
  \doi{https://doi.org/10.1016/j.csda.2018.03.007},
  \urlprefix\url{http://www.sciencedirect.com/science/article/pii/S0167947318300550}

\bibitem[{Papastamoulis and Iliopoulos(2009)}]{papRJ}
Papastamoulis P, Iliopoulos G (2009) Reversible jump {M}{C}{M}{C} in mixtures
  of normal distributions with the same component means. Computational
  Statistics and Data Analysis 53(4):900--911

\bibitem[{Papastamoulis and Iliopoulos(2010)}]{Papastamoulis:10}
Papastamoulis P, Iliopoulos G (2010) An artificial allocations based solution
  to the label switching problem in {B}ayesian analysis of mixtures of
  distributions. Journal of Computational and Graphical Statistics 19:313--331

\bibitem[{Papastamoulis and Iliopoulos(2013)}]{Papastamoulis2013}
Papastamoulis P, Iliopoulos G (2013) On the convergence rate of random
  permutation sampler and {E}{C}{R} algorithm in missing data models.
  Methodology and Computing in Applied Probability 15(2):293--304,
  \doi{10.1007/s11009-011-9238-7},
  \urlprefix\url{http://dx.doi.org/10.1007/s11009-011-9238-7}

\bibitem[{Papastamoulis and Rattray(2017)}]{papastamoulis2016bayesbinmix}
Papastamoulis P, Rattray M (2017) {BayesBinMix: an {R} {P}ackage for {M}odel
  {B}ased {C}lustering of {M}ultivariate {B}inary {D}ata}. {The R Journal}
  9(1):403--420,
  \urlprefix\url{https://journal.r-project.org/archive/2017/RJ-2017-022/index.html}

\bibitem[{Plummer et~al.(2006)Plummer, Best, Cowles, and Vines}]{coda}
Plummer M, Best N, Cowles K, Vines K (2006) {C}{O}{D}{A}: Convergence diagnosis
  and output analysis for {M}{C}{M}{C}. R News 6(1):7--11,
  \urlprefix\url{https://journal.r-project.org/archive/}

\bibitem[{{R Core Team}(2016)}]{R}
{R Core Team} (2016) R: A Language and Environment for Statistical Computing. R
  Foundation for Statistical Computing, Vienna, Austria,
  \urlprefix\url{https://www.R-project.org/}, {ISBN} 3-900051-07-0

\bibitem[{Rand(1971)}]{doi:10.1080/01621459.1971.10482356}
Rand WM (1971) Objective criteria for the evaluation of clustering methods.
  Journal of the American Statistical Association 66(336):846--850

\bibitem[{Redner and Walker(1984)}]{redner1984mixture}
Redner RA, Walker HF (1984) Mixture densities, maximum likelihood and the
  {E}{M} algorithm. SIAM review 26(2):195--239

\bibitem[{{Revolution Analytics and Steve Weston}(2014)}]{foreach}
{Revolution Analytics and Steve Weston} (2014) foreach: Foreach looping
  construct for R. \urlprefix\url{http://CRAN.R-project.org/package=foreach}, r
  package version 1.4.2

\bibitem[{{Revolution Analytics and Steve Weston}(2015)}]{doparallel}
{Revolution Analytics and Steve Weston} (2015) doParallel: Foreach Parallel
  Adaptor for the 'parallel' Package.
  \urlprefix\url{http://CRAN.R-project.org/package=doParallel}, r package
  version 1.0.10

\bibitem[{Richardson and Green(1997)}]{Richardson:97}
Richardson S, Green PJ (1997) On {B}ayesian analysis of mixtures with an
  unknown number of components. Journal of the Royal Statistical Society:
  Series B 59(4):731--758

\bibitem[{Rousseau and Mengersen(2011)}]{rousseau2011asymptotic}
Rousseau J, Mengersen K (2011) Asymptotic behaviour of the posterior
  distribution in overfitted mixture models. Journal of the Royal Statistical
  Society: Series B (Statistical Methodology) 73(5):689--710

\bibitem[{Schwarz(1978)}]{Schwarz:78}
Schwarz G (1978) Estimating the dimension of a model. The Annals of Statistics
  6(2):461--464

\bibitem[{Stephens(2000)}]{stephens2000}
Stephens M (2000) {B}ayesian analysis of mixture models with an unknown number
  of components -- an alternative to reversible jump methods. Annals of
  Statistics 28(1):40--74

\bibitem[{Streuli(1973)}]{Streuli}
Streuli H (1973) Der heutige stand der kaffeechemie. In: 6th International
  Colloquium on Coffee Chemisrty, Association Scientifique International du
  Cafe, Bogata, Columbia, pp 61--72

\bibitem[{Tipping and Bishop(1999)}]{tipping1999mixtures}
Tipping ME, Bishop CM (1999) Mixtures of probabilistic principal component
  analyzers. Neural computation 11(2):443--482

\bibitem[{Yeung et~al.(2001)Yeung, Fraley, Murua, Raftery, and
  Ruzzo}]{doi:10.1093/bioinformatics/17.10.977}
Yeung KY, Fraley C, Murua A, Raftery AE, Ruzzo WL (2001) Model-based clustering
  and data transformations for gene expression data. Bioinformatics
  17(10):977--987, \doi{10.1093/bioinformatics/17.10.977},
  \urlprefix\url{http://dx.doi.org/10.1093/bioinformatics/17.10.977},
  \eprint{/oup/backfile/content_public/journal/bioinformatics/17/10/10.1093/bioinformatics/17.10.977/2/170977.pdf}

\end{thebibliography}

\end{document}